\documentclass[pra,aps,superscriptaddress,twocolumn]{revtex4-2}

\usepackage{amsmath}
\usepackage{amssymb}
\usepackage{physics}
\usepackage{graphicx}
\usepackage[colorlinks=true,linkcolor=blue,citecolor=red,urlcolor=magenta]{hyperref}
\usepackage{xcolor}
\usepackage{bbold}
\newtheorem{theorem}{Theorem}
\newtheorem{proposition}[theorem]{Proposition}

\newtheorem{definition}{Definition}

\usepackage{adjustbox}

\begin{document}

\title{Pauli Flat Quantum States Mimicking Maximal Magic}

\begin{abstract}
Quantum magic is a necessary resource for fault-tolerant quantum computing, capturing the departure of quantum states from the classically simulable stabiliser set. Motivated by its role in quantum information processing, we investigate the correlation structure of states that maximize the R{\'e}nyi stabiliser entropies, focusing on the T-type and Hoggar states. Inspired by their distinctive properties, we explore the notion of \emph{Pauli flat states}, defined as $N$-qubit states whose expectation values on all Pauli strings are equal in absolute value. These states reproduce the correlation structure of highly magical states, which could also be interpreted as a mixed variant of Pauli covariant $N$-qubit symmetric informationally complete (SIC) Weyl-Heisenberg fiducial states, while not necessarily possessing nontrivial quantum magic. We investigate their existence and possible constructions, including examples both inside and outside of the stabiliser polytope. To identify the latter, we examine a magic witness tailored to the introduced family of states. Finally, we discuss their entanglement and Bell nonlocality, providing a broader characterization of their quantum properties. For completeness, we find pure states with a relaxed condition on the Pauli expectations, which forces unequal elements to zero, and discuss their non-stabiliserness. Our results offer a new perspective on the relation between quantum magic and Pauli correlation structures. They hold relevance to geometrical approaches to quantum mechanics, entanglement and magic detection, as well as estimation tasks.

\end{abstract}

\author{Pawe{\l} Cie{\'s}li{\'n}ski}
\email{pawel.cieslinski@nus.edu.sg}
\affiliation{Centre for Quantum Technologies, National University of Singapore, Singapore 117543, Singapore}

\maketitle

\section{Introduction}

Fault-tolerant quantum computing offers a significant impact on science and technology. Despite many uncertainties, examining the resources responsible for computational improvement beyond purely classical machines is also of great foundational importance. Throughout the years, among others, the source of quantum advantage was ascribed to two of the most archetypal quantum phenomena, i.e., superposition and entanglement; see~\cite{Deutsch_1985} and~\cite{Jozsa_2003} for early references. The most recent understanding of it comes from the Gottesman-Knill theorem~\cite{Gottesman_1997, Gottesman_1998}. It shows that even highly coherent and entangled states can be efficiently simulated on classical computers, and the 'magic' behind quantum computing needs operations on states that lie outside of the stabiliser set. Hence, the non-stabiliserness, or quantum magic, is a feature of quantum systems that is a necessary prerequisite for quantum computing.

A primary focus of quantum magic research concerns quantum algorithms, fault tolerance, and complexity, see e.g.~\cite{Aaronson_2004, Bravyi_2005, Campbell_2010, Bravyi_2016, Bravyi_2016b, Seddon_2021, Haug_2026, Leone_2026}. However, recently, an increasing amount of attention has been given to magic as a resource itself, and its impact on the structure of quantum states. New measures and witnesses of quantum magic are being proposed~\cite{Campbell_2011, Mari_2012, Veitch_2014, Howard_2017RoM, Wang_2020, Leone_2022, Haug_2023}, including  characterisations of the stabiliser polytope~\cite{Palhares_2026, Leone_2026}. Moreover, its relation to entanglement and Bell nonlocality is vastly examined~\cite{Howard_2012, Howard_2014, Howard_2015, Tirrito_2024, Bejan_2024, Iannotti_2025, Gu_2025, Macedo_2025, Cusumano_2025, Cieslinski_2026, Viscardi_2026}, together with its ability to explore many-body systems behaviour~\cite{Liu_2022, Fu_2022, Tarabunga_2024, Cieslinski_2026_spins}.  
Despite a big effort to understand the nature of non-stabiliserness, a lot of it remains to be uncovered. Only recently, it was shown that the maximally magical pure states with respect to a computable measure of magic called stabiliser Renyi entropy (SRE)~\cite{Leone_2022} are necessarily Weil-Heisenberg symmetric informationally complete projective operator-valued measure (WH-SIC POVM) fiducial states~\cite{Cuffaro_2024}. This conclusion is conditioned on the WH-SIC existence, which for an arbitrary dimension is still conjectured~\cite{Appleby_2005, Scott_2010, Fuchs_2017}. For the case of the qubit states, which have the biggest computational relevance, such maximal magic pure states exist only for one and three qubits~\cite{Godsil_2009}. The best two-qubit state achieving maximal SRE was constructed recently by Liu, Low, and Yin in~\cite{Liu_2026_2qmagic}. Although known from the mutually unbiased bases (MUB) research, it came from an explicit SRE maximisation.\\
\indent Here we propose and explore the notion of \emph{Pauli flat states}, which resemble the correlation structure of the maximally non-stabiliser states for one and three qubits in a multipartite setting. We define them as states with a flat distribution on \emph{all the Pauli strings' expectation values} (up to a sign) and show their existence for any number of parties. Furthermore, we provide non-stabiliserness witnesses and construct Pauli flat states which lie outside of the stabiliser polytope. We discuss sufficient conditions for their entanglement and verify their Bell nonlocal behaviour. Our findings show that mixed multipartite qubit states can closely mimic the maximal magic of pure states and provide a new correlation structure unique to quantum mechanics. They offer a deeper insight into understanding of the operational manifestations of  quantum magic available from the Pauli measurements and its interplay with other nonclassical resources.

\section{Motivation} \label{sec:motivation}

We start by examining expectation values of Pauli strings for pure states that maximise quantum magic as measured via SRE. Let $P_i=\langle \sigma_{i_1} \otimes \cdots \otimes \sigma_{i_N} \rangle$, where $i$ serves as a multi-index and $\sigma_{i_k}=\{ \openone, X, Y, Z\}$ are the Pauli matrices. The ordering of terms is dictated by the tensor product of local Paulis, i.e. $\{P_i\}_{i=0}^{4^N}=\{\openone^{\otimes N}, \openone^{\otimes N-1} X, \openone^{\otimes N-1} Y, \cdots, Z^{\otimes N} \}$. Magic measure of our interest, the stabiliser $2$-R{\'e}nyi entropy, for an $N$-qubit pure state $|\psi \rangle$ is defined as~\cite{Leone_2022}
\begin{equation}
    M_2(|\psi\rangle)=-\log \frac{1}{2^N} \sum_i P_i^4,
\end{equation}
where the sum goes over all possible Pauli strings. For pure stabiliser states $P_i=\{0,\pm 1\}$ and $M_2$ simply vanishes. The above can be seen as the entropy of a distribution of the normalised set of $P_i^2$ values. Hence, one expects that $N$-qubit states containing the highest amount of magic would have (if possible) all of the Pauli expectations equal to $\pm \gamma$, where, as discussed at the end of this section, its purity requires $\gamma=\sqrt{1/(2^N+1)}$. The set of absolute values of all Pauli expectations is often referred to as a Pauli spectrum. Indeed, such a structure is present in the maximally magical states.  
In the one-qubit case, the state of interest is given by the so-called T-type magic state (also referred to as a golden state)~\cite{Bravyi_2005}
\begin{equation}
    |T\rangle= \sqrt{\frac{1+1/\sqrt{3}}{2}}|0\rangle + e^{i \pi/4}\sqrt{\frac{1-1/\sqrt{3}}{2}}|1\rangle.
\end{equation}
Its Bloch decomposition reads
\begin{equation}
    \rho_{\,T}=\frac{1}{2}\left( \openone + \frac{1}{\sqrt{3}}\left[X + Y +Z \right]\right).
\end{equation}
The above state has the highest SRE out of all single-qubit states, and all of its Pauli expectations are non-zero and yield $1/\sqrt{3}$. For three qubits, the same property is present in the Hoggar state~\cite{Hoggar_1998}
\begin{eqnarray} \label{eq:Hoggar_state}
    |\mathrm{Hoggar}\rangle &=&\frac{1}{\sqrt{6}}[(1+i)|000\rangle - |010\rangle \nonumber \\
    &+& |011\rangle -i|100\rangle + |101\rangle],
\end{eqnarray}
which again contains the maximal magic in the sense discussed before, and all of its Pauli expectations are non-zero and equal in absolute value, i.e., all 63 of its Pauli string expectations follow
\begin{equation}
    \forall P_i \neq \openone^{\otimes N}:  \Tr (\rho_{|{\rm Hoggar}\rangle} P_i)=\pm \frac{1}{3}.
\end{equation}

The reason for omitting the $N=2$ case is that, for $N=2$ and $N>3$, no such state can be constructed~\cite{Godsil_2009}. This follows from the fact that any such state $|\psi\rangle$ (necessarily not being an eigenstate of any $P_i$) has to fulfill $|\langle P_i\rangle |^2=|\langle \psi | P_i |\psi\rangle |^2=1/(2^N+1)$ for all non-identity $i$. The left-hand side, however, is equivalent to $|\langle \psi | (X^{p_1}Z^{q_1}) \otimes \cdots  \otimes (X^{p_N}Z^{q_N})|\psi\rangle |^2$, for all $(\Vec{p},\Vec{q})\neq \vec{0}$,
which for $d=2^N$ is the condition for an existence of a WH-SIC fiducial state of $N$-qubit Weyl-Heisenberg group. This only exists for $N=1,3$~\cite{Godsil_2009}, which makes such a construction with a purity constraint impossible.

For illustration, we perform an explicit proof for $N=2$ using the Pauli expectation values (here often referred to as correlations) language.
For a pure state $\rho=|\psi\rangle \langle \psi |$ to have all Pauli expectations equal up to a sign, say as before $P_i=\pm \gamma$, one needs to have $\Tr \rho^2=[1+\gamma^2 (4^2-1)]/2^2=1$, from which $\gamma=\pm1/\sqrt{5}$ follows. Assume such a state exists. We can express it in the Schmidt form with the coefficients parametrised by an angle $\theta$, i.e., $\lambda_1=\cos\theta$ and $\lambda_2=\sin \theta$. Then, a $3\times3$ matrix $\mathbb{T}$ composed of Pauli expectations supported on two qubits is diagonal with $T_{11}=\sin 2\theta$, $T_{22}=-\sin 2\theta$, and $T_{33}=1$. This matrix forms the $2$-qubit block of the so-called correlation tensor. The local Bloch vector components of the Schmidt form state are given as $\cos 2\theta$ in the $Z$ direction only. The $\mathbb{T}$ matrix of the Schmidt form state is equivalent to that of the conjectured state up to local orthogonal transformations~\cite{Horodecki_1996}. This means that $\mathrm{Det}(\mathbb{T})$ should be equal to multiples of $1/5\sqrt{5}$ as this is the case for the conjectured state. However, from the invariance of local Bloch vectors' lengths, $\theta$ should be such that $\cos^2 2\theta=3 \cdot (1/\sqrt{5})^2=3/5$. For an angle $\theta$ fulfilling this condition, it also has to hold that $|\mathrm{Det}(\mathbb{T})|=\sin^2 2\theta=2/5$, which \emph{is not a multiple} of $1/5\sqrt{5}$, and hence a two-qubit pure state with all correlations equal in absolute value cannot be constructed. 

\section{Pauli Flat States}
\label{sec:main}

Based on the impossibility of constructing pure states with equal absolute values on all Pauli strings beyond the one- and three-qubit cases, we move on to mixed states and ask how well you can mimic such a structure in this setting. We will refer to such states as \emph{Pauli flat}.
However, as we are interested in states whose correlations are as different from $\pm 1$ as possible, while not being $0$, one could also name them inverse stabiliser states. For completeness, we provide the definition below.
\begin{definition}
An $N$-qubit state satisfying
\begin{equation}
    \Tr [\rho P_i]=\pm \, \gamma, \quad \forall P_i\neq \openone^{\otimes N},
\end{equation}
where $P_i$ stands for tensor products of Pauli matrices, is called a Pauli flat state. For a sign vector $\Vec{r}$ with $\pm 1$ entries, they can be expressed as
\begin{equation}
    \rho = \frac{1}{2^N}(\openone^{\otimes N}+\gamma \sum_{i> 0}  r_i  P_i ).
\end{equation}
\end{definition}

From the reasoning presented in the previous section, \emph{pure Pauli flat states exist only for $N=1$ and $N=3$ qubits}. Before we move to the exploration of mixed Pauli flat states, we discuss alternative ways of approaching the problem of interest. Namely, one may wonder why the purity condition is relaxed in our definition. Another way of looking for states with a similar structure could be to define a probability distribution associated with the Pauli strings expectations via $q_i=1/2^N \langle P_i \rangle^2$ and try to optimise it towards the desired flatness. As the uniform discrete probability distributions are the ones that maximise the entropy, it could become the object of interest. However, by taking the entropy to be of the $\alpha$-Renyi type, this approach corresponds exactly to the reasoning from~\cite{Leone_2022} that introduced SRE. Maximisation of SRE then yields the results shown in~\cite{Cuffaro_2024}. Since the desired solution exists only for one and three qubits, our definition is well justified. Another approach to the problem would be to examine approximate pure Pauli flat states. Reasonable definitions could impose the equality of Pauli expectations of a given weight, a set of weights, or another specific subset of Pauli strings. This is the case for the two-qubit state of maximal SRE magic constructed by Liu, Low, and Yin (LLY)~\cite{Liu_2026_2qmagic}. It is given as
\begin{equation}
    |\mathrm{LLY}\rangle=\frac{1}{2}\left[i(|00\rangle+|01\rangle +|10\rangle) + |11\rangle \right] .
    \label{eq:LLY}
\end{equation}
All of its Pauli expectation values except for $Z \openone, \openone  Z$ and $Z Z$ are equal to $\pm 1/2$. The above state is an MUB fiducial for two qubits, and was also shown to maximise a closely related nonstabiliserness quantifier called magick~\cite{Damski_2026}. Following the LLY conjecture on maximal magic $N$-qubit states given by either WH-SIC or WH-MUB (conditioned on their existence), identifying such states as approximate Pauli flat states would lead to an increasingly sparser non-zero correlation structure.  Although very exciting, this direction is of independent interest. Nevertheless, numerical results for pure states achieving the highest found number of non-zero Paulis, with the rest being equal in absolute value, are presented in Sec.~\ref{sec:pure_states_relax} for completeness. 

Given the above justification of our state definition, we continue the exploration of the mixed Pauli flat states by formulating the following Proposition.
\begin{proposition}
    Mixed Pauli flat states exist for an arbitrary number of qubits $N$. Moreover, there exists such $N\neq 1,3$ that they cannot be expressed as a convex mixture of pure stabiliser states.
\end{proposition}
The existence of Pauli flat states for all $N$ can be proved by finding such $\gamma$ that any sign combination in $\Vec{r}$ leads to a proper (positive semi-definite) density matrix. Pauli flat states can be interpreted geometrically as high-dimensional vectors with equal norm of their projection onto all possible axes. Since beyond one qubit there is no Bloch sphere, it is rather intuitive that these vectors need to be shrunk in order to exhibit such behaviour while still generating a positive density matrix. In fact, it is fascinating that this can be done for $N=3$ with a pure state. A much harder task is to find such Pauli flat states that do not lie within the stabiliser polytope. We prove the first statement here and show examples of non-zero magic Pauli flat states by explicit construction afterwards. 

For $\rho$ to be a proper density matrix, one has to have $ \langle \rho \rangle \geq 0$ for all $|\psi\rangle$. The correlation term can be bounded as
\begin{eqnarray}
    \langle \rho \rangle =\frac{1}{2^N} (\openone^{\otimes N}+\gamma \sum_{i >0} r_i \langle P_i \rangle) \geq \frac{1}{2^N} (\openone^{\otimes N}-\gamma |\sum_{i>0} r_i \langle P_i \rangle|) \nonumber \\
    \geq \frac{1}{2^N} (\openone^{\otimes N}-\gamma \sqrt{(4^N-1)(2^N-1)}) \nonumber
\end{eqnarray}
where the last follows from the Cauchy-Schwarz inequality. The sign vector $\Vec{r}$ has $4^N-1$ entries with square norm of the same value, and the correlation term has to respect the upper bound on purity $(\sum_i \langle P_i \rangle ^2)/2^N \leq 1$ from which the expression under the square root follows.  Then, one gets that the above is non-negative for
\begin{equation}
    \gamma \leq \frac{1}{\sqrt{(4^N -1)(2^N-1)}}
\end{equation}
Thus, for any $N$, choosing $\gamma$ in compliance with the above yields a proper state which is also Pauli flat. For now we omit the non-trivial magic statement, as it will be proven by an explicit construction. By taking an algebraic lower bound on the spectrum of the Pauli term, which is more rigorous, one can obtain $\gamma \leq 1/(4^N-1)$. Every Pauli flat state respecting it necessarily lies within the stabiliser polytope. This is clear, as one can simply construct an equally weighted mixture of a single stabiliser mixed states that reconstructs it exactly. Although the bound is not optimal, the difference in these ranges of $\gamma$ hints that a magic Pauli flat state construction is possible. 

\section{Tailored magic witness}

A condition on $\gamma$ that signals non-zero magic can be derived via a witness approach. Here, we define the following operator
\begin{equation} \label{eq:W}
    \mathcal{W}=\sum_{i>0} \Tilde{r}_i P_i, \quad \Tilde{r}_i=\pm1,
\end{equation}
which, by construction, achieves its maximal expectation value on a Pauli flat state when all $\Tilde{r}_i=r_i$. Then
\begin{equation}
    \Tr(\rho \, \mathcal{W})=\sum_i |\gamma|=\gamma (4^N-1).
    \label{eq:witness_on_ISS}
\end{equation}
For any stabiliser state, we have
\begin{eqnarray}
    \Tr(\rho_S \, \mathcal{W})=\sum_k p_k \Tr( |S\rangle \langle S| \, \mathcal{W}) \leq \max_S \Tr( |S\rangle \langle S| \, \mathcal{W}) \nonumber,
\end{eqnarray}
where $|S\rangle$ denotes a pure stabiliser state (graph state)~\cite{Briegel_2001, Russendorf_2001, Hein_2004}. A set of all such mixtures forms a stabiliser polytope denoted as $\mathrm{Conv}(S)=\mathrm{STAB}$. For any $|S\rangle$, one can have at most $2^N-1$ non-zero Pauli expectations. Thus, even by neglecting the possible loss of positivity due to an improper placement of $\pm 1$ values onto the Pauli string (improper choice of $\Vec{r}$), it has to hold that
\begin{eqnarray}
    \Tr(\rho_S \, \mathcal{W}) \leq \max_S \Tr( |S\rangle \langle S| \, \mathcal{W}) \nonumber \\
    = \sum_{i \in \mathrm{supp}(S)} r_i \langle S |P_i |S\rangle \leq 2^N-1, 
\end{eqnarray}
where $\mathrm{supp}(S)$ denotes the set of non-zero correlations in $|S\rangle$. Inspired by geometric entanglement witnesses (see e.g.~\cite{Toth_2005, Toth_2007}), we formulate the following criterion.
\begin{proposition}
For any $N$-qubit state $\rho$ and $\mathcal{W}$ as defined in~(\ref{eq:W}), it holds that
\begin{eqnarray}
    \text{if} \quad \Tr(\rho \, \mathcal{W})> 2^N-1 \Rightarrow \rho \notin {\rm STAB}.
\end{eqnarray}   
\end{proposition}
Now, comparing (\ref{eq:witness_on_ISS}) with the above we get that
 \begin{equation}
     \forall \, \gamma > \frac{1}{2^N+1}:  \quad \rho \notin {\rm STAB},
 \end{equation}
conditioned on a proper choice of $\Vec{r}$ such that $\rho \succeq 0$.
The above witness can detect weak magic in Pauli flat states, which are just outside of the STAB. The stabiliser condition for $\gamma$ is equivalent to the positivity of the mixed states efficient magic witnesses from~\cite{Haug_2026} for $\alpha=1/2$ and reduced robustness of magic for a full set of Pauli strings~\cite{Varela_2026}. The former follows from the fact that the witness value for Pauli flat states reduces to the stabiliser norm witness~\cite{Campbell_2011}.

\section{Explicit few-qubit Pauli flat states}

Consider the following family of two-qubit states parametrised by $\gamma$
 \begin{equation}
     \rho^{(2)}=\begin{bmatrix}
         \frac{1}{4}(1-3\gamma) & 0 & 0 & 0\\
         0 & \frac{1+\gamma}{4} & (\frac{1}{2}+\frac{i}{2})\gamma & -(\frac{1}{2}+\frac{i}{2})\gamma \\
         0 & (\frac{1}{2}-\frac{i}{2})\gamma & \frac{1+\gamma}{4} & -(\frac{1}{2}+\frac{i}{2})\gamma\\
         0 & (-\frac{1}{2}+\frac{i}{2})\gamma & (-\frac{1}{2}+\frac{i}{2})\gamma & \frac{1+\gamma}{4}
     \end{bmatrix} \nonumber
 \end{equation}
One can easily verify that its Pauli expectations are 
\begin{eqnarray}
    \langle X \openone \rangle = \langle  \openone  X\rangle = \langle   Z \openone\rangle =   \langle  \openone  Z\rangle= \langle Y  X\rangle \nonumber\\
    =\langle YZ \rangle= \langle ZY \rangle=\langle ZZ \rangle=-\gamma, \nonumber \\
     \langle Y \openone \rangle = \langle  \openone  Y\rangle =\langle X  X\rangle =\langle X  Y \rangle=\langle XZ\rangle \nonumber \\
     =\langle YY \rangle=\langle ZX \rangle=\gamma, \nonumber
\end{eqnarray}
For $N=2$, the stabiliser bound on $\gamma$ is given as $1/5$. We find that the above state is positive for $\gamma \leq 1/3$. Taking the highest possible value and setting admissible sign vector as
\begin{eqnarray*}
    \ \vec{r}&=&\{1, -1,+1,-1,-1,+1,+1,+1,+1,\\
    &&-1,+1, -1, -1,+1,-1,-1 \}
\end{eqnarray*}
yields a proper Pauli flat state which is not in STAB, and hence contains magic.  In fact, such a statement holds for all $\gamma \in (1/5,1/3]$. For the highest identified correlation value, it is equivalent to a reduced state of (\ref{eq:Hoggar_state}) under local Clifford operations. The considered state is entangled, as verified by the PPT criterion. For $\gamma=1/3$ it has a negativity~\cite{Zyczkowski_1998, Vidal_2002} of $\approx-0.18$, which corresponds to the minimal eigenvalue of $\rho^{(2)}$ after a partial transposition. Another interesting question about its resource-theoretic nature would concern Bell nonlocality. Despite being entangled, however, it is not Bell nonlocal in the two-output two-setting scenario. This can be shown directly from the $\mathbb{T}$ matrix defined in Sec.~\ref{sec:motivation} using the singular values criterion from ~\cite{Horodecki_1995}. The singular values of $\mathbb{T}$ for the considered state are given as $(2/3,2/3,1/3)$. Based on~\cite{Horodecki_1995}, a two-qubit state violates the Clauser-Horne-Shimony-Holtz (CHSH) inequality iff the sum of the two highest squares of singular values of $\mathbb{T}$ is greater than 1. For the examined state, one has $8/9<1$ and the maximal CHSH value reads $4\sqrt{2}/3\approx 1.88 <2$.
\begin{figure}
    \centering
    \includegraphics[width=1\linewidth]{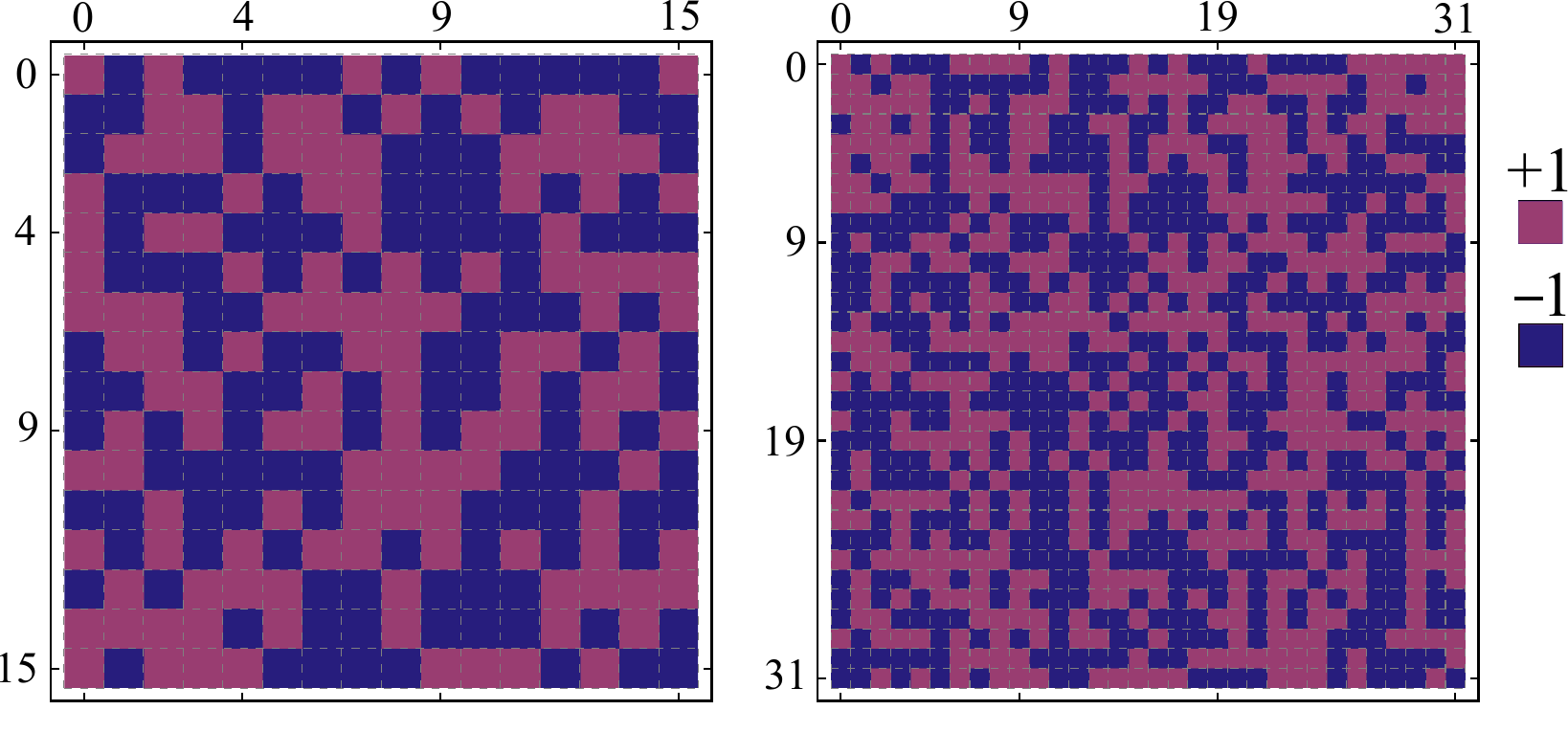}
    \caption{Matrix representation of the sign vectors $\Vec{r}$ for best-found Pauli flat states of $N=4$ and $N=5$ qubits. Each matrix is constructed by reshaping $\Vec{r}$ into a $2^N$ dimensional array with rows corresponding to its consecutive $2^N$-tuples. }
    \label{fig:sign_matrix}
\end{figure}

A general way of finding such states can be done via semidefinite programming (SDP). We are interested in Pauli flat states, which have the highest correlations and hence the highest purity. Since the purity of such a state is a monotonic function of $\gamma$, for a given sign vector, one can maximize $\gamma$ under positivity constraints. Explicitly, we have 
\begin{eqnarray}
    \max&& \gamma\\
    s.t.&& \rho^{(N)}_{\vec{r}}=\frac{1}{2^N} (\openone^{\otimes N}+\gamma \sum_{i>0} r_i  P_i )  \succeq 0  \nonumber
\end{eqnarray}
Repeating the above over different choices of $\vec{r}$ or optimizing over them directly allows one to find an optimal solution. Due to the sign vector domain size, this is the most daunting task in the routine. 
An alternative formulation of the above problem concerns eigenvalue (ground state energy) optimisation. By global sign change, the highest excited state and the ground state of the Pauli Hamiltonian $H=\sum_{i>0} r_i P_i=r_1 X_N + r_2 Y_N + \ldots + r_{4^N}Z_1 \cdots Z_N$ can be swapped. Thus, finding the optimal $\gamma$ is equivalent to finding $\max_{\vec{r}} 1/\lambda_{\max}(H)$.
Here, we provide the best-found $\gamma$ for $N=2,4,5$ in Table~\ref{tab:tab}. They are complemented by the non-stabiliserness witness ratio $\eta_S=\Tr(\rho \, \mathcal{W})/(2^N-1)$ that signals $\rho$ not being a member of the stabiliser polytope whenever it is $>1$. Since even for the non-existing case of pure states, $\gamma$ needs to decrease with $N$, we also present their respective purities. This captures how close the presented Pauli flat states are to the impossible-to-construct pure state cases, and how much one has to shrink the Bloch vector in order to satisfy the uniform projection onto all axes condition. 

\begin{table}[ht!] \vspace{0.2cm}
\centering
\begin{tabular}{|c|c|c|c|c|c|}
\hline
$N$ & 1 & 2 & 3 & 4 & 5  \\
\hline \hline
$\gamma$ & $ 0.5773$ & $ 0.3334$ & $0.3334$ & $0.0673$ & $0.0293$  \\
\hline
Purity & $1$ & $0.6667$ & $1$ & $0.1347$ & $0.0586$ \\
\hline
$\eta_S$ & $1.73$ & $1.6667$ & $3$ & $1.1442$ & $0.966$  \\
\hline
\end{tabular}
\caption{Best-found values of $\gamma$ for Pauli flat states of $N=2,4,5$ qubits, together with the state's purity and magic witness ratio $\eta_S$. Whenever $\eta_S>1$, the underlying state is not a stabiliser. Explicit values for the T-type ($N=1$) and Hoggar ($N=3$) states are also provided. Despite analytical values for $N<4$, the numerical form was provided for clearer comparison.}
\label{tab:tab}
\end{table}

The sign vectors for the best-found $N=4,5$ cases are provided in a matrix representation in Fig.~\ref{fig:sign_matrix}. 
Together with $\gamma$ values shown in Table~\ref{tab:tab}, they allow an explicit reconstruction of the desired states up to $N\leq 5$ qubits. Explicit sign vectors and parametrised states are provided in Appendix~\ref{app} for convenience.

For $N=4$, the best $\gamma$ was found to be $\approx 0.067$ with $\eta_S \approx 1.14$, which signals non-stabiliserness in the state. In contrast to the Pauli flat state of two qubits, it cannot be generated by tracing out ancillary parties from a larger pure Pauli flat state. Its entanglement can again be inferred from the PPT criterion. The considered state is definitely entangled as, for each bipartition, the sum of all negative eigenvalues ranges between $0.066$ and $0.096$. Despite that, it is either not, or very weakly, genuinely multipartite entangled. This was verified by the monotone from~\cite{Jungnitsch_2011} using the PPTMixer tool~\cite{bastian2026pptmixer}.
A four-partite state Bell nonlocality is significantly harder to examine. However, limiting oneself to two and three settings per party, it becomes feasible. An especially interesting class of inequalities concerning Pauli flat states are Bell inequalities that can be violated with Pauli measurements only.  Using techniques from~\cite{Cieslinski_2026} it can be verified that, despite having all non-zero correlations, it does not lead to a violation of any two- and three-settings Bell inequality of this type.

In the last example of $N=5$ an appropriate Pauli flat state was found, but the corresponding $\gamma$ was too small to yield any magic within it. Due to its correlation structure and no detectable magic, we regard its operational meaning as magic mimicking. The same behaviour is present in all previous states if $\gamma$ is diminished below the stabiliser bound. Despite the purity of $\approx 0.06$, the five-qubit Pauli flat state is still entangled, and its negativity over all bipartitions yields values within $[0.05,0.066]$, which has a non-zero overlap with the values obtained for $N=4$. Using the same techniques as before, the considered state is concluded to have no genuine multipartite entanglement.

\subsection{Relaxation on the number of equal correlations}
\label{sec:pure_states_relax}

\begin{figure}
    \centering
    \includegraphics[width=0.85\linewidth]{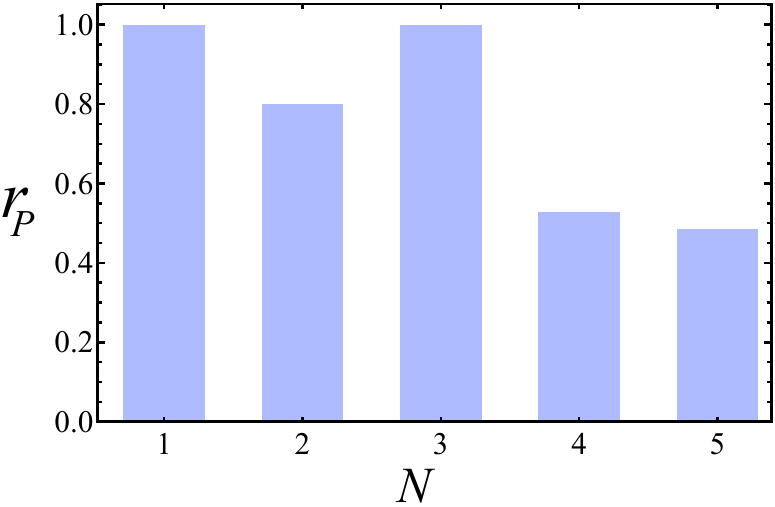}
    \caption{Ratio $r_P$ for pure states with an approximate Pauli flat state structure of up to five qubits. Top bar values correspond to the number of non-zero and equal in absolute value Pauli expectations divided by the number of all non-identity Pauli strings. Due to their definition, all other expectations are strictly 0, and $r_P$ has an interpretation of correlation density.}
    \label{fig:rp}
\end{figure}

For completeness, we also examine \emph{pure} states which do not have \emph{all} of the Pauli expectations equal in absolute value. For extended discussion see Sec.\ref{sec:main}. In order to have a structure that resembles Pauli flat states closely, we follow the LLY state example. We thus look for a state that contains $m$ Pauli expectations equal up to a sign, with the rest of them being 0. For (\ref{eq:LLY}) $m=12$ and the remaining $(4^2-1)-12=3$ correlations are 0. 

As discussed before, constructing such states is closely related to the product WH group MUB and SIC fiducials, which lie beyond the scope of this work. Here, we present analytical states inferred from the numerical search for completeness only. Our objective function to be minimised over multiple choices of $\gamma^*$ is
\begin{equation}
    f(|\psi\rangle)=\sum_i \langle P_i \rangle_{\psi}^2 \left[\langle P_i \rangle_{\psi}^2-\frac{1}{(\gamma^{*})^2}\right].
\end{equation}
The above achieves 0 when all Pauli expectations are 0 or $\pm 1/\gamma^*$. Let us define the ratio of non-zero to all Pauli expectations as $r_P$. Since the state is pure and the remaining Paulis need to have 0 expectation value by definition, the target value of $\gamma^*$ in terms of $r_p$ is an inverse of $\pm \sqrt{r_p(2^N+1)}$. Hence, it acts as a shrinking factor.

For one qubit, the T-type state is recovered and $r_p=1$. For the two-qubit case, the same correlations as in the LLY state were found. For $N=3$, as should be, the Hoggar state structure was obtained. For $N=4$ and $5$ we found the following\\
\begin{eqnarray}
    |s_4\rangle&=&\frac{1}{\sqrt{24}}(-3,1,1,1, \cdots ,1,1,1),\\
    |s_5\rangle &=& \frac{1}{8 \sqrt{5}} (\alpha_1,\,-\alpha_2,\,\alpha_1,\,-\alpha_2,\,-\alpha_1,\,-\alpha_2,\,-\alpha_1,\, \nonumber\\
    && \alpha_2, i\alpha_3,\,\alpha_1,\,\alpha_2,\,\alpha_3,\,\alpha_2,\,\alpha_4,\,-\alpha_2,\,\alpha_3, \, \alpha_2, \nonumber\\
    && \alpha_3,\,-\alpha_2,\,-\alpha_1,\,\alpha_2,\,-\alpha_4,\,-\alpha_2,\,-\alpha_1, \, \alpha_1, \nonumber\\
    && -i\alpha_3,\,-\alpha_1,\,\alpha_2,\,-\alpha_3,\,\alpha_2,\,-\alpha_1,\,-\alpha_2),
\end{eqnarray}
where $\alpha_1 = -1-2i, \alpha_2 = 2-i, \alpha_3 = 5,\alpha_4 = -3+4i$. The non-zero equal Pauli expectations $m$ are 135 and 496, respectively. Ratios $r_P$ for $N\leq 5$ are shown in Fig.~\ref{fig:rp}.
Interestingly, the first three states were proven to maximise SRE. The newly found, or recovered (as we do not claim to be the first one presenting them), states are also far from being a stabiliser. The numerically found approximate Pauli flat states yield SRE which is reasonably close to the allowed maximum. For $|s_4\rangle$ the SRE reads $1.79$, while the maximum according to Cuffaro and Fuchs~\cite{Cuffaro_2024} is $2.14$, and due to LLY conjecture~\cite{Liu_2026_2qmagic} $2.11$. For $|s_5 \rangle$ one gets $2.39$ with respective bounds of $2.8$ and $2.79$. 

\section{Conclusions}

Inspired by the correlation structure of T-type and Hoggar states, we introduced the notion of Pauli flat states. These were defined as states (possibly mixed) that have all of the Pauli strings' expectation values equal in absolute value.  We showed their existence for an arbitrary number of qubits and provided examples for which such states are not stabilisers. This analysis was possible due to the magic witness tailored specifically to the states of interest.  Other resources, such as entanglement and Bell nonlocality, were also discussed. Complementary calculations for an approximate structure of this type, where the number of equal correlations (with the non-equal ones being zero) is traded for the state's purity, were also performed and recovered SIC fiducials for $N=1,3$ and MUB fiducial for $N=2$. For $N=4,5$ states with 135 and 496 equal in absolute value Pauli expectations and high stabiliser R{\'e}nyi entropy values were found. 

Pauli flat states seem interesting from multiple different perspectives. Geometrically, they describe a high-dimensional vector whose projection onto any axis is equal in modulus. Informationally, or from the pure mathematics point of view, they could be regarded as a mixed version of $N$-qubit Pauli group covariant WH-SIC fiducials. This suggests that they may find applications in, among others, tomography protocols; however, here fundamentally limited by the presence of noise. Pauli flat states could also gain attention from the entanglement theory community. Many experimentally friendly criteria for entanglement are based on the sector lengths~\cite{Hassan_2008, Badziag_2008, Tran_2015, Ketterer_2019, Wyderka_2020}, which are easily characterised here. In the past, examples of entangled states with no full-body correlations~\cite{Tran_2017} or the least amount of correlations were discussed. The states discussed here provide the opposite family of states. Finally, Pauli flat states show that high quantum magic (as captured via SRE) could be mimicked by stabiliser states. Not measuring all of the Pauli strings (which allow one to compute the state's purity) could lead to interpreting the high entropy of $\langle P_i \rangle^2$ distribution as associated with a magical state. Adaptive measurement protocols (see e.g.~\cite{Laskowski_2013}) based on the commutation structure of the Pauli group  together with the examined witness can serve as new means of efficient magic certification~\cite{Varela_2026}. More distant perspectives may include Hamiltonian learning usefulness based on the non-zero overlap with all terms in an arbitrary Pauli Hamiltonian, or metrology, where no specific "correlation axis" might become a resource in multi-parameter estimation tasks.

Future work on the introduced concepts includes finding the optimal Pauli flat states, examining their entanglement and Bell nonlocal properties more thoroughly (especially in the context of XYZ Bell inequalities~\cite{Cieslinski_2026}), and exploring their performance in magic distillation and injection protocols. Stronger connections to mixed $t$-designs, MUBs, weighing matrices and SIC-POVMs are also worth investigating. 

\section*{Acknowledgements}
We thank Otfried G{\"u}hne, Wies{\l}aw Laskowski, Mengyao Hu, and Dagomir Kaszlikowski for stimulating discussions. Moreover, we express our gratitude to Markus Grassl and Jakub Czartowski for insightful comments on WH-SICs, MUBs, and related concepts. This project is supported by the National Research Foundation, Singapore through the National Quantum Office, hosted in A*STAR, under its Centre for Quantum Technologies Funding Initiative (S24Q2d0009). \\

\textit{AI usage:} AI tools were used for language checking and text formatting (ChatGPT, Grammarly), code optimisation (ChatGPT, Claude), and technical proofreading after the manuscript's completion (ChatGPT). For numerical calculations, an initial code used for preliminary results was fed to ChatGPT and Claude for optimisation and improvement. Later, the outputs were verified by an independent non-AI-written code.

\bibliographystyle{apsrev4-2}
\bibliography{ref}

\appendix
\onecolumngrid
\newpage
\section{Explicit sign vectors and states}
\label{app}

Here we present explicit sign vectors and density matrices for Pauli flat states of $N=4$ and $N=5$ qubits. The one- and three-qubit cases are known in the literature, while the two-qubit one is shown explicitly in the main text. Ordering of the Pauli expectations follows from the tensor product of local Pauli sets, i.e., for $N=4$ we have
\begin{eqnarray*}
    \{ \openone ^{\otimes 3} X, \openone ^{\otimes 3} Y, \openone ^{\otimes 3} Z, \openone ^{\otimes 2} X \openone, \openone ^{\otimes 2} XX, \cdots, Z ^{\otimes 3} Y, Z ^{\otimes 4} \}.
\end{eqnarray*}
The sign vector $\Vec{r}$ for the best found $\gamma=0.06730677027165417$ reads
\begin{eqnarray*}
&&\{-, +, -, -, -, -, +, -, +, -, -, -, -, -, +, -, -, +, +,
-, +, +, -, +, -, +, -, +, +, -, -, -, +, +, +, -, +, +, \\
&& +,-, -, -, +, +, +, +, -, +, -, -, -, +, -, +, +, -, -, -, 
+, -, +, -, +, +, -, +, +, -, -, -, +, -, -, -, -, +,  \\
&&-,-, -, +, -, -, -, +, -, +, -, +, -, +, -, +, +, +, +, +, +, 
+, -, -, +, +, +, +, +, -, -, -, +, -, +, -, +, +, -,  \\
&& +,-,-, +, +, -, -, +, +, -, +, -, -, -, +, +, -, -, +, -, +, 
-, -, +, -, +, +, -, -, +, -, +, -, +, +, -, +, -, +,\\
&& +,  -,+, -, +, +, +, -, -, -, -, -, +, +, +, +, -, -, -, +, -, 
-, -, +, -, -, +, -, +, +, +, -, -, -, +, -, -, +, -, \\
&& +,-, +, -, +, +, -, +, -, +, -, +, -, +, -, +, -, +, +, +, -, 
-, +, -, -, -, +, +, +, +, +, +, +, +, -, +, -, -, +, \\
&& -, -,-, +, -, +, -, +, -, +, +, +, -, -, -, -, +, +, +, -, +, -, 
-\}
\end{eqnarray*}

The explicit density matrix diagonal elements are
\begin{align*}
\rho_{0,0}&=\frac{1}{16}(3\gamma+1), & \rho_{1,1}&=\frac{1-\gamma}{16}, & \rho_{2,2}&=\frac{1-\gamma}{16}, & \rho_{3,3}&=\frac{1}{16}(3\gamma+1),\\
\rho_{4,4}&=\frac{1}{16}(3\gamma+1), & \rho_{5,5}&=\frac{1-\gamma}{16}, & \rho_{6,6}&=\frac{1-\gamma}{16}, & \rho_{7,7}&=\frac{1}{16}(3\gamma+1),\\
\rho_{8,8}&=\frac{1}{16}(3\gamma+1), & \rho_{9,9}&=\frac{1}{16}(7\gamma+1), & \rho_{10,10}&=\frac{1}{16}(1-9\gamma), & \rho_{11,11}&=\frac{1}{16}(3\gamma+1),\\
\rho_{12,12}&=\frac{1}{16}(1-5\gamma), & \rho_{13,13}&=\frac{1-\gamma}{16}, & \rho_{14,14}&=\frac{1-\gamma}{16}, & \rho_{15,15}&=\frac{1}{16}(1-5\gamma).
\end{align*}

The off-diagonal (upper-triangular) elements are given as
\begin{align*}
    &\rho_{0,1}=\frac{3i\gamma}{8},\ \rho_{0,2}=-\frac{\gamma}{4},\ \rho_{0,3}=\left(\frac38-\frac i4\right)\gamma,\ \rho_{0,4}=\left(-\frac18+\frac i4\right)\gamma,\ \rho_{0,5}=\frac{i\gamma}{8},\ \rho_{0,6}=0,\\
&\rho_{0,7}=\frac{i\gamma}{8},\ \rho_{0,8}=\left(\frac18+\frac i4\right)\gamma,\ \rho_{0,9}=-\frac{\gamma}{8},\ \rho_{0,10}=\frac{3\gamma}{8},\ \rho_{0,11}=\frac{\gamma}{4},\ \rho_{0,12}=0,\\
&\rho_{0,13}=\left(\frac14+\frac i4\right)\gamma,\ \rho_{0,14}=\left(-\frac38+\frac i8\right)\gamma,\ \rho_{0,15}=\left(-\frac18+\frac i4\right)\gamma,\\
&\rho_{1,2}=-\frac{\gamma}{8},\ \rho_{1,3}=-\frac{i\gamma}{4},\ \rho_{1,4}=-\frac{i\gamma}{8},\ \rho_{1,5}=\frac{\gamma}{8},\ \rho_{1,6}=\frac{i\gamma}{8},\ \rho_{1,7}=\left(\frac12-\frac i4\right)\gamma,\\
&\rho_{1,8}=-\frac{\gamma}{8},\ \rho_{1,9}=\frac{\gamma}{8},\ \rho_{1,10}=0,\ \rho_{1,11}=\frac{\gamma}{8},\ \rho_{1,12}=0,\ \rho_{1,13}=-\frac{i\gamma}{4},\\
&\rho_{1,14}=\left(\frac18-\frac i4\right)\gamma,\ \rho_{1,15}=\left(-\frac18-\frac i8\right)\gamma,\\
\end{align*}
\begin{align*}
&\rho_{2,3}=\left(-\frac14-\frac i8\right)\gamma,\ \rho_{2,4}=0,\ \rho_{2,5}=\left(\frac12+\frac i8\right)\gamma,\ \rho_{2,6}=\frac{\gamma}{8},\ \rho_{2,7}=\frac{i\gamma}{8},\\
&\rho_{2,8}=\left(\frac18-\frac i4\right)\gamma,\ \rho_{2,9}=\frac{i\gamma}{2},\ \rho_{2,10}=\frac{\gamma}{8},\ \rho_{2,11}=\left(-\frac18-\frac i4\right)\gamma,\ \rho_{2,12}=\left(\frac18+\frac i8\right)\gamma,\\
&\rho_{2,13}=-\frac{\gamma}{8},\ \rho_{2,14}=-\frac{i\gamma}{4},\ \rho_{2,15}=-\frac{i\gamma}{4},\\
&\rho_{3,4}=\frac{i\gamma}{8},\ \rho_{3,5}=-\frac{i\gamma}{4},\ \rho_{3,6}=\frac{3i\gamma}{8},\ \rho_{3,7}=\left(-\frac18-\frac i4\right)\gamma,\ \rho_{3,8}=\frac{\gamma}{4},\\
&\rho_{3,9}=\left(-\frac18-\frac i4\right)\gamma,\ \rho_{3,10}=\left(-\frac18+\frac i4\right)\gamma,\ \rho_{3,11}=\left(\frac18+\frac i4\right)\gamma,\ \rho_{3,12}=\frac{\gamma}{8},\\
&\rho_{3,13}=\left(-\frac18-\frac i8\right)\gamma,\ \rho_{3,14}=-\frac{\gamma}{4},\ \rho_{3,15}=0,\\
&\rho_{4,5}=-\frac{i\gamma}{8},\ \rho_{4,6}=\left(-\frac14-\frac i4\right)\gamma,\ \rho_{4,7}=\frac{\gamma}{8},\ \rho_{4,8}=\left(-\frac14+\frac i4\right)\gamma,\ \rho_{4,9}=0,\\
&\rho_{4,10}=\left(-\frac18+\frac i8\right)\gamma,\ \rho_{4,11}=\left(\frac18-\frac i4\right)\gamma,\ \rho_{4,12}=\left(-\frac18-\frac i4\right)\gamma,\ \rho_{4,13}=\frac{\gamma}{8},\\
&\rho_{4,14}=\left(\frac18+\frac i4\right)\gamma,\ \rho_{4,15}=0,\\
&\rho_{5,6}=\left(\frac18+\frac i4\right)\gamma,\ \rho_{5,7}=0,\ \rho_{5,8}=\left(\frac14+\frac i4\right)\gamma,\ \rho_{5,9}=\frac{\gamma}{4},\ \rho_{5,10}=\left(-\frac18-\frac i4\right)\gamma,\\
&\rho_{5,11}=\left(\frac18-\frac i8\right)\gamma,\ \rho_{5,12}=-\frac{3\gamma}{8},\ \rho_{5,13}=-\frac{\gamma}{8},\ \rho_{5,14}=\frac{\gamma}{4},\ \rho_{5,15}=\left(-\frac18-\frac i4\right)\gamma,\\
&\rho_{6,7}=\left(-\frac14-\frac i8\right)\gamma,\ \rho_{6,8}=\left(-\frac18+\frac i8\right)\gamma,\ \rho_{6,9}=-\frac{3\gamma}{8},\ \rho_{6,10}=-\frac{\gamma}{4},\ \rho_{6,11}=\frac{\gamma}{4},\\
&\rho_{6,12}=-\frac{\gamma}{8},\ \rho_{6,13}=-\frac{\gamma}{4},\ \rho_{6,14}=-\frac{\gamma}{8},\ \rho_{6,15}=\left(\frac18-\frac i4\right)\gamma,\\
&\rho_{7,8}=-\frac{\gamma}{8},\ \rho_{7,9}=\left(\frac18+\frac{3i}{8}\right)\gamma,\ \rho_{7,10}=\frac{i\gamma}{4},\ \rho_{7,11}=\left(-\frac14-\frac i4\right)\gamma,\ \rho_{7,12}=0,\\
&\rho_{7,13}=\frac{\gamma}{8},\ \rho_{7,14}=\left(\frac18-\frac i4\right)\gamma,\ \rho_{7,15}=\left(\frac38+\frac i4\right)\gamma,\\
&\rho_{8,9}=-\frac{i\gamma}{8},\ \rho_{8,10}=\left(\frac14+\frac i4\right)\gamma,\ \rho_{8,11}=\frac{\gamma}{8},\ \rho_{8,12}=\frac{\gamma}{8},\ \rho_{8,13}=\left(-\frac14-\frac i8\right)\gamma,\\
&\rho_{8,14}=\frac{i\gamma}{4},\ \rho_{8,15}=-\frac{i\gamma}{8},\\
&\rho_{9,10}=\left(\frac18+\frac i4\right)\gamma,\ \rho_{9,11}=0,\ \rho_{9,12}=\left(\frac14+\frac i8\right)\gamma,\ \rho_{9,13}=\left(-\frac18-\frac i4\right)\gamma,\\
&\rho_{9,14}=-\frac{i\gamma}{8},\ \rho_{9,15}=0,\\
&\rho_{10,11}=\left(-\frac14-\frac i8\right)\gamma,\ \rho_{10,12}=-\frac{i\gamma}{4},\ \rho_{10,13}=\frac{3i\gamma}{8},\ \rho_{10,14}=\left(-\frac18-\frac i4\right)\gamma,\ \rho_{10,15}=\left(\frac14-\frac i8\right)\gamma,\\
&\rho_{11,12}=-\frac{i\gamma}{8},\ \rho_{11,13}=0,\ \rho_{11,14}=\left(\frac14+\frac i8\right)\gamma,\ \rho_{11,15}=-\frac{3\gamma}{8},\\
&\rho_{12,13}=-\frac{i\gamma}{8},\ \rho_{12,14}=-\frac{\gamma}{4},\ \rho_{12,15}=\left(-\frac18+\frac i4\right)\gamma,\\
&\rho_{13,14}=-\frac{\gamma}{8},\ \rho_{13,15}=-\frac{i\gamma}{4},\qquad \rho_{14,15}=\left(\frac14-\frac i8\right)\gamma.
\end{align*}
The full matrix is reconstructed via $\rho_{ji}=\rho_{ij}^{*}$ for all $j>i$. \\

For $N=5$ and the best found $\gamma=0.029272337634361104$, the sign vector reads
\begin{small}
\begin{eqnarray*}
&&\{
-, +, -, -, -, +, +, +, +, -, +, -, -, -, +, -, +, -, -, 
-, +, -, -, -, -, +, +, +, +, +, +, +, +, -, +, +, -, -, -, \\
&&-, -, -, +, -, -, +, +, +, +, +, -, -, -, +, +, +, +, -, +, 
+, -, +, +, +, +, +, +, +, -, -, +, -, +, +, +, -, -, -, +, \\
&&-, +, -, -, +, +, -, -, +, -, -, +, +, +, +, +, -, +, +, -, 
+, -, +, -, -, +, +, -, -, +, +, -, -, +, -, +, +, +, +, -, \\
&&+, -, +, +, -, +, +, +, +, +, +, +, +, -, +, -, -, +, +, -, -, 
-, +, -, +, +, +, -, -, +, +, -, +, +, -, +, -, -, -, -, +, \\
&&-, +, +, -, -, +, +, -, +, -, -, -, -, +, -, +, -, +, +, 
-, +, +, +, -, +, -, -, +, +, -, -, +, +, -, +, +, -, +, +, \\
&&+, +, +, +, +, -, +, +, -, -, -, +, -, +, +, -, -, -, -, -, 
-, -, +, +, +, +, +, -, -, -, -, +, -, +, +, +, +, -, +, -, \\
&&-, -, -, +, +, +, +, +, +, -, -, +, -, +, -, +, -, -, -, 
-, -, -, +, -, +, -, -, -, +, -, +, -, -, -, -, -, +, \\
&&-, +, -, -, -, +, -, -, -, -, +, -, +, -, -, +, +, -, +, 
+, -, -, +, -, -, -, +, -, +, -, +, -, +, +, +, -, +, +, -, \\
&&+, +, +, -, -, -, +, +, -, +, +, -, -, +, +, +, -, -, -, -, 
+, +, -, +, +, -, -, +, +, +, +, +, -, -, -, -, -, -, +, -, \\
&&-, -, -, +, -, -, +, -, +, -, -, +, -, +, +, +, -, -, +, 
-, +, -, +, -, -, +, -, +, -, -, +, -, +, -, -, +, +, -, \\
&&-, +, +, -, +, -, +, -, +, -, -, +, -, -, -, -, -, +, +, 
+, +, +, -, +, -, -, -, +, -, +, -, +, +, +, +, +, -, +, +, +, \\
&&+, +, -, +, +, +, -, +, -, +, +, -, +, -, +, +, +, -, -, +, +, 
+, +, +, +, +, -, +, +, -, -, +, -, +, -, -, -, +, -, -, +, \\
&&-, +, +, -, -, -, +, +, +, -, -, -, -, +, -, +, +, -, -, 
-, +, -, -, +, -, +, +, -, +, +, +, +, +, +, +, -, +, +, -, +, \\
&&-, +, +, +, +, -, -, -, -, +, -, +, -, -, +, -, +, -, +, 
-, +, +, -, +, +, -, -, -, +, -, -, -, -, -, -, +, -, -, \\
&&-, -, -, -, +, -, +, -, -, +, +, -, -, -, +, +, -, +, +, 
-, +, -, -, +, -, -, +, -, -, +, +, +, -, -, -, +, -, +, \\
&&-, -, +, -, +, -, -, +, +, +, +, -, -, +, +, +, +, -, -, -, 
+, +, +, +, +, -, +, -, -, +, -, -, -, +, -, -, +, +, -, -, \\
&&+, +, +, +, +, -, +, -, +, -, +, -, -, -, +, -, +, -, +, -, 
+, -, +, -, -, +, -, -, +, -, -, +, -, +, -, -, +, +, -, \\
&&+, -, -, +, -, -, -, -, +, -, +, -, -, -, +, -, +, +, +, 
+, -, -, -, +, +, +, +, -, -, -, -, +, -, +, +, -, +, +, +, \\
&&+, -, +, -, +, -, -, +, -, +, +, +, +, +, +, +, -, -, +, -, 
+, -, +, -, +, -, -, +, +, -, +, -, -, -, +, -, +, -, -, \\
&&+, -, +, +, -, +, -, +, -, +, -, +, -, +, +, +, -, +, -, +, 
-, -, -, +, -, +, -, -, +, -, -, -, +, -, +, -, -, -, +, \\
&&+, +, +, -, +, +, -, -, -, -, +, -, +, +, -, +, +, +, +, +, +, 
+, -, -, -, -, +, -, -, -, -, -, +, -, -, -, -, +, -, \\
&&+, -, -, +, +, +, -, +, -, -, +, +, -, +, -, +, +, -, -, +, 
+, +, +, -, -, -, +, -, +, +, -, +, +, -, -, +, -, +, -, +, \\
&&-, +, -, +, +, +, -, +, -, -, -, +, +, -, +, -, +, -, +, -, 
+, -, +, -, +, -, -, +, -, -, -, +, +, -, -, -, -, +, +, \\
&&+, +, -, +, -, -, +, -, -, -, +, +, -, +, -, +, +, -, -, 
-, +, -, -, +, -, +, +, +, -, +, -, +, -, +, -, +, +, -, -, \\
&&+, -, -, -, +, -, +, +, +, -, -, -, +, +, +, +, -, -, -, 
-, -, -, +, +, +, +, -, -, -, -, -, +, -, -, +, +, +, +, +, \\
&&+, +, -, +, -, -, -, -, +, -, -, +, -, +, -, +, -, +, +, +, 
-, -, +, +, +, +, +, -, -, -, -, +, +, +, -, +, -, -, -, +, 
-\}
\end{eqnarray*}
\end{small}
The diagonal elaments are given as
\begin{align*}
\rho_{0,0}&=\frac{\gamma +1}{32},\
\rho_{1,1}=\frac{1}{32} (1-7 \gamma ),\
\rho_{2,2}=\frac{1}{32} (1-3 \gamma ),\
\rho_{3,3}=\frac{1}{32} (5 \gamma +1),\
\rho_{4,4}=\frac{\gamma +1}{32},\
\rho_{5,5}=\frac{\gamma +1}{32}\\
\rho_{6,6}&=\frac{1}{32} (1-3 \gamma ),\
\rho_{7,7}=\frac{1}{32} (1-3 \gamma ),\
\rho_{8,8}=\frac{\gamma +1}{32},\
\rho_{9,9}=\frac{\gamma +1}{32},\
\rho_{10,10}=\frac{1}{32} (1-3 \gamma ),\
\rho_{11,11}=\frac{1}{32} (1-3 \gamma )\\
\rho_{12,12}&=\frac{1}{32} (1-7 \gamma ),\
\rho_{13,13}=\frac{\gamma +1}{32},\
\rho_{14,14}=\frac{1}{32} (5 \gamma +1),\
\rho_{15,15}=\frac{1}{32} (1-3 \gamma ),\
\rho_{16,16}=\frac{1}{32} (9 \gamma +1),\
\rho_{17,17}=\frac{\gamma +1}{32}\\
\rho_{18,18}&=\frac{1}{32} (1-3 \gamma ),\
\rho_{19,19}=\frac{1}{32} (5 \gamma +1),\
\rho_{20,20}=\frac{\gamma +1}{32},\
\rho_{21,21}=\frac{\gamma +1}{32},\
\rho_{22,22}=\frac{1}{32} (5 \gamma +1),\
\rho_{23,23}=\frac{1}{32} (5 \gamma +1)\\
\rho_{24,24}&=\frac{1}{32} (1-7 \gamma ),\
\rho_{25,25}=\frac{1}{32} (9 \gamma +1),\
\rho_{26,26}=\frac{1}{32} (1-3 \gamma ),\
\rho_{27,27}=\frac{1}{32} (13 \gamma +1),\
\rho_{28,28}=\frac{1}{32} (1-7 \gamma )\\
\rho_{29,29}&=\frac{1}{32} (1-15 \gamma ),\
\rho_{30,30}=\frac{1}{32} (1-3 \gamma ),\
\rho_{31,31}=\frac{1}{32} (5 \gamma +1)
\end{align*}
The off-diagonal elements are
\begin{align*}
\rho_{0,1}&=\left(-\frac{3}{16}+\frac{i}{16}\right) \gamma,\
\rho_{0,2}=-\frac{1}{4} (i \gamma ),\
\rho_{0,3}=\left(-\frac{1}{16}-\frac{i}{8}\right) \gamma,\
\rho_{0,4}=0,\
\rho_{0,5}=-\frac{1}{16} (3 \gamma )\\
\rho_{0,6}&=0,\
\rho_{0,7}=\left(\frac{1}{8}+\frac{3 i}{16}\right) \gamma,\
\rho_{0,8}=\left(-\frac{1}{16}-\frac{i}{8}\right) \gamma,\
\rho_{0,9}=\frac{\gamma }{16},\
\rho_{0,10}=\left(\frac{1}{4}-\frac{i}{16}\right) \gamma\\
\rho_{0,11}&=\left(-\frac{1}{4}+\frac{i}{16}\right) \gamma,\
\rho_{0,12}=\left(-\frac{1}{16}+\frac{5 i}{16}\right) \gamma,\
\rho_{0,13}=\frac{\gamma }{8},\
\rho_{0,14}=\frac{\gamma }{8},\
\rho_{0,15}=-\frac{1}{16} (3 \gamma )\\
\rho_{0,16}&=\left(-\frac{1}{8}-\frac{i}{16}\right) \gamma,\
\rho_{0,17}=0,\
\rho_{0,18}=\left(\frac{1}{16}-\frac{3 i}{16}\right) \gamma,\
\rho_{0,19}=\left(\frac{1}{16}-\frac{3 i}{16}\right) \gamma,\
\rho_{0,20}=\left(-\frac{1}{8}+\frac{i}{16}\right) \gamma
\end{align*}
\begin{align*}
\rho_{0,21}&=\frac{\gamma }{8},\
\rho_{0,22}=\left(-\frac{1}{8}-\frac{i}{16}\right) \gamma,\
\rho_{0,23}=\left(\frac{1}{16}+\frac{i}{16}\right) \gamma,\
\rho_{0,24}=\left(\frac{1}{16}-\frac{i}{8}\right) \gamma,\
\rho_{0,25}=-\frac{1}{16} (3 \gamma )\\
\rho_{0,26}&=\left(-\frac{1}{4}-\frac{i}{8}\right) \gamma,\
\rho_{0,27}=\left(-\frac{3}{16}-\frac{i}{16}\right) \gamma,\
\rho_{0,28}=\left(-\frac{1}{16}-\frac{i}{8}\right) \gamma,\
\rho_{0,29}=-\frac{1}{16} (i \gamma ),\
\rho_{0,30}=\left(\frac{1}{8}-\frac{i}{16}\right) \gamma\\
\rho_{0,31}&=\left(-\frac{1}{16}+\frac{i}{16}\right) \gamma,\
\rho_{1,2}=\left(-\frac{3}{16}+\frac{i}{8}\right) \gamma,\
\rho_{1,3}=\left(-\frac{1}{4}-\frac{i}{4}\right) \gamma,\
\rho_{1,4}=\frac{\gamma }{16},\
\rho_{1,5}=-\frac{1}{8} (3 \gamma )\\
\rho_{1,6}&=\left(\frac{1}{8}+\frac{3 i}{16}\right) \gamma,\
\rho_{1,7}=-\frac{1}{8} (i \gamma ),\
\rho_{1,8}=\frac{3 \gamma }{16},\
\rho_{1,9}=\left(-\frac{3}{16}-\frac{i}{8}\right) \gamma,\
\rho_{1,10}=\left(\frac{1}{8}+\frac{i}{16}\right) \gamma\\
\rho_{1,11}&=\frac{i \gamma }{16},\
\rho_{1,12}=\left(\frac{1}{8}-\frac{i}{4}\right) \gamma,\
\rho_{1,13}=\left(-\frac{1}{16}+\frac{i}{16}\right) \gamma,\
\rho_{1,14}=\frac{\gamma }{16},\
\rho_{1,15}=0\\
\rho_{1,16}&=\left(\frac{1}{8}+\frac{i}{8}\right) \gamma,\
\rho_{1,17}=\left(\frac{1}{8}+\frac{i}{16}\right) \gamma,\
\rho_{1,18}=\left(\frac{1}{16}-\frac{i}{16}\right) \gamma,\
\rho_{1,19}=\left(-\frac{1}{16}+\frac{i}{16}\right) \gamma,\
\rho_{1,20}=\left(-\frac{1}{8}-\frac{i}{8}\right) \gamma\\
\rho_{1,21}&=\frac{5 i \gamma }{16},\
\rho_{1,22}=\left(\frac{1}{16}+\frac{i}{16}\right) \gamma,\
\rho_{1,23}=\left(\frac{1}{8}+\frac{i}{16}\right) \gamma,\
\rho_{1,24}=\left(-\frac{1}{16}+\frac{i}{4}\right) \gamma,\
\rho_{1,25}=\left(-\frac{3}{16}-\frac{i}{8}\right) \gamma\\
\rho_{1,26}&=\left(\frac{1}{16}+\frac{3 i}{16}\right) \gamma,\
\rho_{1,27}=\left(\frac{1}{8}+\frac{i}{8}\right) \gamma,\
\rho_{1,28}=\left(\frac{1}{8}+\frac{i}{16}\right) \gamma,\
\rho_{1,29}=-\frac{\gamma }{16},\
\rho_{1,30}=\left(-\frac{1}{16}-\frac{i}{16}\right) \gamma\\
\rho_{1,31}&=\left(\frac{1}{8}+\frac{i}{16}\right) \gamma,\
\rho_{2,3}=\left(\frac{1}{16}+\frac{3 i}{16}\right) \gamma,\
\rho_{2,4}=-\frac{1}{8} (i \gamma ),\
\rho_{2,5}=\left(\frac{1}{8}-\frac{i}{16}\right) \gamma,\
\rho_{2,6}=\left(-\frac{1}{8}-\frac{i}{8}\right) \gamma\\
\rho_{2,7}&=-\frac{\gamma }{16},\
\rho_{2,8}=-\frac{1}{16} (i \gamma ),\
\rho_{2,9}=\left(\frac{1}{8}+\frac{i}{16}\right) \gamma,\
\rho_{2,10}=\frac{\gamma }{16},\
\rho_{2,11}=-\frac{\gamma }{16}\\
\rho_{2,12}&=\frac{i \gamma }{8},\
\rho_{2,13}=\left(\frac{1}{16}+\frac{i}{8}\right) \gamma,\
\rho_{2,14}=\left(-\frac{5}{16}-\frac{3 i}{16}\right) \gamma,\
\rho_{2,15}=\frac{i \gamma }{8},\
\rho_{2,16}=\left(-\frac{1}{16}-\frac{i}{16}\right) \gamma\\
\rho_{2,17}&=\left(\frac{3}{16}-\frac{i}{16}\right) \gamma,\
\rho_{2,18}=\left(-\frac{1}{8}-\frac{i}{16}\right) \gamma,\
\rho_{2,19}=\left(-\frac{1}{8}+\frac{i}{8}\right) \gamma,\
\rho_{2,20}=\left(\frac{1}{4}-\frac{i}{16}\right) \gamma,\
\rho_{2,21}=\left(\frac{3}{16}-\frac{i}{16}\right) \gamma\\
\rho_{2,22}&=-\frac{1}{16} (i \gamma ),\
\rho_{2,23}=\left(-\frac{1}{8}-\frac{i}{4}\right) \gamma,\
\rho_{2,24}=-\frac{1}{8} (i \gamma ),\
\rho_{2,25}=\left(\frac{1}{16}+\frac{i}{16}\right) \gamma,\
\rho_{2,26}=\left(\frac{1}{16}-\frac{i}{8}\right) \gamma\\
\rho_{2,27}&=\left(\frac{1}{16}-\frac{i}{8}\right) \gamma,\
\rho_{2,28}=\frac{i \gamma }{16},\
\rho_{2,29}=\left(\frac{1}{16}+\frac{i}{16}\right) \gamma,\
\rho_{2,30}=\left(\frac{3}{16}+\frac{i}{8}\right) \gamma,\
\rho_{2,31}=\frac{i \gamma }{16}\\
\rho_{3,4}&=\left(\frac{1}{8}-\frac{i}{16}\right) \gamma,\
\rho_{3,5}=0,\
\rho_{3,6}=-\frac{\gamma }{16},\
\rho_{3,7}=-\frac{1}{8} (i \gamma ),\
\rho_{3,8}=\frac{i \gamma }{16}\\
\rho_{3,9}&=\left(\frac{1}{4}+\frac{i}{16}\right) \gamma,\
\rho_{3,10}=\frac{\gamma }{16},\
\rho_{3,11}=\frac{3 \gamma }{16},\
\rho_{3,12}=\left(\frac{5}{16}-\frac{i}{8}\right) \gamma,\
\rho_{3,13}=\left(\frac{1}{8}-\frac{i}{8}\right) \gamma\\
\rho_{3,14}&=-\frac{1}{8} (i \gamma ),\
\rho_{3,15}=\left(\frac{3}{16}+\frac{i}{16}\right) \gamma,\
\rho_{3,16}=\left(-\frac{1}{16}+\frac{i}{16}\right) \gamma,\
\rho_{3,17}=\left(-\frac{3}{16}+\frac{3 i}{16}\right) \gamma,\
\rho_{3,18}=0\\
\rho_{3,19}&=\left(-\frac{1}{8}+\frac{i}{16}\right) \gamma,\
\rho_{3,20}=\left(-\frac{5}{16}-\frac{i}{16}\right) \gamma,\
\rho_{3,21}=\frac{i \gamma }{16},\
\rho_{3,22}=\left(\frac{1}{8}-\frac{i}{8}\right) \gamma,\
\rho_{3,23}=\left(-\frac{1}{8}-\frac{i}{16}\right) \gamma\\
\rho_{3,24}&=\left(-\frac{3}{16}+\frac{i}{16}\right) \gamma,\
\rho_{3,25}=\left(\frac{1}{8}+\frac{i}{8}\right) \gamma,\
\rho_{3,26}=\left(-\frac{1}{16}+\frac{i}{8}\right) \gamma,\
\rho_{3,27}=\left(-\frac{3}{16}+\frac{i}{8}\right) \gamma,\
\rho_{3,28}=\left(-\frac{3}{16}-\frac{i}{16}\right) \gamma\\
\rho_{3,29}&=\frac{3 i \gamma }{16},\
\rho_{3,30}=\left(\frac{1}{8}+\frac{3 i}{16}\right) \gamma,\
\rho_{3,31}=\frac{3 \gamma }{16},\
\rho_{4,5}=\left(-\frac{3}{16}+\frac{i}{16}\right) \gamma,\
\rho_{4,6}=0\\
\rho_{4,7}&=\left(\frac{1}{16}-\frac{i}{8}\right) \gamma,\
\rho_{4,8}=\left(\frac{1}{16}-\frac{i}{16}\right) \gamma,\
\rho_{4,9}=\left(\frac{1}{8}+\frac{i}{8}\right) \gamma,\
\rho_{4,10}=\left(-\frac{1}{8}+\frac{i}{8}\right) \gamma,\
\rho_{4,11}=\left(-\frac{1}{16}-\frac{i}{8}\right) \gamma\\
\rho_{4,12}&=\left(\frac{1}{16}+\frac{i}{8}\right) \gamma,\
\rho_{4,13}=-\frac{\gamma }{16},\
\rho_{4,14}=\left(\frac{1}{8}+\frac{3 i}{16}\right) \gamma,\
\rho_{4,15}=\frac{3 i \gamma }{16},\
\rho_{4,16}=-\frac{1}{16} (i \gamma )\\
\rho_{4,17}&=\frac{i \gamma }{8},\
\rho_{4,18}=\frac{3 i \gamma }{16},\
\rho_{4,19}=\left(-\frac{3}{16}+\frac{i}{16}\right) \gamma,\
\rho_{4,20}=\left(-\frac{1}{8}-\frac{3 i}{16}\right) \gamma,\
\rho_{4,21}=\left(-\frac{1}{4}+\frac{i}{8}\right) \gamma\\
\rho_{4,22}&=\left(\frac{1}{16}-\frac{3 i}{16}\right) \gamma,\
\rho_{4,23}=\left(-\frac{1}{16}-\frac{3 i}{16}\right) \gamma,\
\rho_{4,24}=\left(-\frac{1}{16}-\frac{i}{8}\right) \gamma,\
\rho_{4,25}=\frac{i \gamma }{16},\
\rho_{4,26}=\left(\frac{1}{8}+\frac{3 i}{16}\right) \gamma\\
\rho_{4,27}&=\left(-\frac{1}{16}-\frac{3 i}{16}\right) \gamma,\
\rho_{4,28}=\left(\frac{1}{16}+\frac{i}{8}\right) \gamma,\
\rho_{4,29}=\left(-\frac{3}{16}+\frac{i}{8}\right) \gamma,\
\rho_{4,30}=0,\
\rho_{4,31}=\left(\frac{3}{16}-\frac{i}{16}\right) \gamma
\end{align*}
\begin{align*}
\rho_{5,6}&=\left(-\frac{1}{16}-\frac{i}{8}\right) \gamma,\
\rho_{5,7}=0,\
\rho_{5,8}=\left(-\frac{1}{8}+\frac{i}{8}\right) \gamma,\
\rho_{5,9}=\left(\frac{1}{16}-\frac{i}{16}\right) \gamma,\
\rho_{5,10}=\left(-\frac{1}{16}+\frac{i}{8}\right) \gamma\\
\rho_{5,11}&=-\frac{1}{8} (i \gamma ),\
\rho_{5,12}=\frac{\gamma }{16},\
\rho_{5,13}=\left(-\frac{1}{16}+\frac{i}{8}\right) \gamma,\
\rho_{5,14}=\left(-\frac{1}{8}-\frac{5 i}{16}\right) \gamma,\
\rho_{5,15}=\left(-\frac{1}{8}+\frac{i}{16}\right) \gamma\\
\rho_{5,16}&=-\frac{1}{4} (i \gamma ),\
\rho_{5,17}=\left(-\frac{1}{8}-\frac{i}{16}\right) \gamma,\
\rho_{5,18}=\left(-\frac{3}{16}+\frac{i}{16}\right) \gamma,\
\rho_{5,19}=\frac{i \gamma }{16},\
\rho_{5,20}=\left(\frac{1}{8}+\frac{i}{4}\right) \gamma\\
\rho_{5,21}&=\left(\frac{1}{8}-\frac{5 i}{16}\right) \gamma,\
\rho_{5,22}=\left(-\frac{1}{16}+\frac{3 i}{16}\right) \gamma,\
\rho_{5,23}=\left(-\frac{1}{16}-\frac{3 i}{16}\right) \gamma,\
\rho_{5,24}=\left(\frac{1}{8}-\frac{5 i}{16}\right) \gamma,\
\rho_{5,25}=-\frac{\gamma }{16}\\
\rho_{5,26}&=\left(\frac{3}{16}-\frac{i}{16}\right) \gamma,\
\rho_{5,27}=\left(-\frac{1}{8}+\frac{i}{16}\right) \gamma,\
\rho_{5,28}=\left(-\frac{1}{16}+\frac{i}{8}\right) \gamma,\
\rho_{5,29}=\left(\frac{1}{16}-\frac{i}{8}\right) \gamma,\
\rho_{5,30}=\left(\frac{3}{16}+\frac{3 i}{16}\right) \gamma\\
\rho_{5,31}&=-\frac{\gamma }{8},\
\rho_{6,7}=\left(\frac{1}{16}-\frac{i}{16}\right) \gamma,\
\rho_{6,8}=\frac{\gamma }{4},\
\rho_{6,9}=-\frac{\gamma }{16},\
\rho_{6,10}=\left(\frac{1}{16}-\frac{i}{16}\right) \gamma\\
\rho_{6,11}&=-\frac{\gamma }{4},\
\rho_{6,12}=\left(\frac{1}{8}-\frac{i}{16}\right) \gamma,\
\rho_{6,13}=\left(\frac{1}{8}+\frac{3 i}{16}\right) \gamma,\
\rho_{6,14}=-\frac{\gamma }{16},\
\rho_{6,15}=\left(\frac{1}{16}-\frac{i}{4}\right) \gamma\\
\rho_{6,16}&=\left(-\frac{1}{8}-\frac{5 i}{16}\right) \gamma,\
\rho_{6,17}=\left(-\frac{1}{16}-\frac{i}{16}\right) \gamma,\
\rho_{6,18}=\left(-\frac{1}{8}-\frac{3 i}{16}\right) \gamma,\
\rho_{6,19}=\frac{i \gamma }{8},\
\rho_{6,20}=\left(-\frac{5}{16}-\frac{i}{16}\right) \gamma\\
\rho_{6,21}&=\left(-\frac{3}{16}-\frac{i}{16}\right) \gamma,\
\rho_{6,22}=\left(\frac{1}{8}+\frac{i}{16}\right) \gamma,\
\rho_{6,23}=\frac{\gamma }{8},\
\rho_{6,24}=\left(\frac{1}{4}+\frac{i}{16}\right) \gamma,\
\rho_{6,25}=\left(\frac{1}{16}+\frac{5 i}{16}\right) \gamma\\
\rho_{6,26}&=\left(-\frac{1}{16}+\frac{i}{8}\right) \gamma,\
\rho_{6,27}=\left(\frac{1}{4}-\frac{i}{16}\right) \gamma,\
\rho_{6,28}=0,\
\rho_{6,29}=\left(-\frac{1}{16}+\frac{i}{16}\right) \gamma,\
\rho_{6,30}=\left(\frac{1}{16}+\frac{i}{8}\right) \gamma\\
\rho_{6,31}&=\frac{\gamma }{16},\
\rho_{7,8}=-\frac{\gamma }{16},\
\rho_{7,9}=\frac{\gamma }{8},\
\rho_{7,10}=0,\
\rho_{7,11}=\left(\frac{1}{16}-\frac{i}{16}\right) \gamma\\
\rho_{7,12}&=\frac{3 i \gamma }{16},\
\rho_{7,13}=\left(-\frac{1}{8}-\frac{3 i}{16}\right) \gamma,\
\rho_{7,14}=\left(\frac{3}{16}-\frac{i}{4}\right) \gamma,\
\rho_{7,15}=\frac{\gamma }{16},\
\rho_{7,16}=\left(-\frac{1}{16}-\frac{i}{16}\right) \gamma\\
\rho_{7,17}&=\left(-\frac{1}{8}+\frac{i}{16}\right) \gamma,\
\rho_{7,18}=0,\
\rho_{7,19}=\frac{i \gamma }{16},\
\rho_{7,20}=\left(\frac{1}{16}-\frac{3 i}{16}\right) \gamma,\
\rho_{7,21}=\left(\frac{1}{16}-\frac{i}{16}\right) \gamma\\
\rho_{7,22}&=-\frac{1}{8} (i \gamma ),\
\rho_{7,23}=\left(\frac{1}{8}-\frac{i}{16}\right) \gamma,\
\rho_{7,24}=\left(\frac{1}{16}-\frac{i}{16}\right) \gamma,\
\rho_{7,25}=-\frac{1}{16} (i \gamma ),\
\rho_{7,26}=\left(-\frac{1}{8}+\frac{i}{16}\right) \gamma\\
\rho_{7,27}&=-\frac{\gamma }{16},\
\rho_{7,28}=\left(-\frac{1}{16}+\frac{i}{16}\right) \gamma,\
\rho_{7,29}=\frac{\gamma }{8},\
\rho_{7,30}=-\frac{\gamma }{16},\
\rho_{7,31}=\left(\frac{1}{16}+\frac{i}{8}\right) \gamma\\
\rho_{8,9}&=\left(-\frac{1}{16}-\frac{3 i}{16}\right) \gamma,\
\rho_{8,10}=-\frac{\gamma }{4},\
\rho_{8,11}=\left(\frac{1}{16}+\frac{i}{8}\right) \gamma,\
\rho_{8,12}=-\frac{1}{8} (i \gamma ),\
\rho_{8,13}=\frac{5 \gamma }{16}\\
\rho_{8,14}&=-\frac{1}{8} (i \gamma ),\
\rho_{8,15}=\left(\frac{1}{8}-\frac{i}{16}\right) \gamma,\
\rho_{8,16}=\left(-\frac{3}{16}-\frac{i}{8}\right) \gamma,\
\rho_{8,17}=\left(\frac{5}{16}+\frac{i}{8}\right) \gamma,\
\rho_{8,18}=-\frac{\gamma }{8}\\
\rho_{8,19}&=\left(\frac{1}{16}+\frac{i}{16}\right) \gamma,\
\rho_{8,20}=\frac{\gamma }{16},\
\rho_{8,21}=\left(\frac{1}{8}+\frac{i}{16}\right) \gamma,\
\rho_{8,22}=-\frac{1}{16} (3 i \gamma ),\
\rho_{8,23}=\left(-\frac{3}{16}+\frac{3 i}{16}\right) \gamma\\
\rho_{8,24}&=\frac{3 i \gamma }{16},\
\rho_{8,25}=\left(-\frac{1}{8}+\frac{i}{8}\right) \gamma,\
\rho_{8,26}=\left(-\frac{3}{16}+\frac{3 i}{16}\right) \gamma,\
\rho_{8,27}=\left(\frac{1}{16}+\frac{i}{16}\right) \gamma,\
\rho_{8,28}=\left(\frac{1}{8}+\frac{i}{16}\right) \gamma\\
\rho_{8,29}&=0,\
\rho_{8,30}=-\frac{1}{16} (i \gamma ),\
\rho_{8,31}=\left(\frac{1}{16}-\frac{3 i}{16}\right) \gamma,\
\rho_{9,10}=\left(-\frac{1}{16}-\frac{i}{8}\right) \gamma,\
\rho_{9,11}=0\\
\rho_{9,12}&=\frac{\gamma }{16},\
\rho_{9,13}=\left(\frac{1}{8}+\frac{i}{8}\right) \gamma,\
\rho_{9,14}=\left(\frac{1}{8}-\frac{i}{16}\right) \gamma,\
\rho_{9,15}=0,\
\rho_{9,16}=\left(-\frac{1}{16}-\frac{i}{8}\right) \gamma\\
\rho_{9,17}&=\left(-\frac{3}{16}+\frac{i}{8}\right) \gamma,\
\rho_{9,18}=\left(\frac{1}{16}+\frac{i}{16}\right) \gamma,\
\rho_{9,19}=-\frac{\gamma }{4},\
\rho_{9,20}=-\frac{1}{16} (i \gamma ),\
\rho_{9,21}=\left(\frac{1}{16}+\frac{i}{8}\right) \gamma\\
\rho_{9,22}&=\left(\frac{1}{16}+\frac{i}{16}\right) \gamma,\
\rho_{9,23}=\left(-\frac{1}{4}+\frac{3 i}{16}\right) \gamma,\
\rho_{9,24}=0,\
\rho_{9,25}=-\frac{1}{16} (3 i \gamma ),\
\rho_{9,26}=\left(\frac{1}{16}+\frac{3 i}{16}\right) \gamma\\
\rho_{9,27}&=\left(-\frac{1}{16}-\frac{i}{16}\right) \gamma,\
\rho_{9,28}=\left(-\frac{1}{4}+\frac{i}{8}\right) \gamma,\
\rho_{9,29}=\frac{i \gamma }{16},\
\rho_{9,30}=\left(\frac{1}{16}-\frac{3 i}{16}\right) \gamma,\
\rho_{9,31}=\frac{i \gamma }{16}\\
\rho_{10,11}&=\left(-\frac{1}{16}-\frac{i}{16}\right) \gamma,\
\rho_{10,12}=0,\
\rho_{10,13}=\left(-\frac{1}{8}+\frac{3 i}{16}\right) \gamma,\
\rho_{10,14}=\left(-\frac{1}{8}-\frac{i}{4}\right) \gamma,\
\rho_{10,15}=-\frac{\gamma }{16}
\end{align*}
\begin{align*}
\rho_{10,16}&=-\frac{\gamma }{8},\
\rho_{10,17}=\left(\frac{1}{16}-\frac{i}{16}\right) \gamma,\
\rho_{10,18}=\left(\frac{1}{16}-\frac{i}{8}\right) \gamma,\
\rho_{10,19}=\frac{\gamma }{16},\
\rho_{10,20}=\left(-\frac{1}{8}+\frac{3 i}{16}\right) \gamma\\
\rho_{10,21}&=\left(-\frac{1}{16}+\frac{3 i}{16}\right) \gamma,\
\rho_{10,22}=\left(\frac{1}{16}+\frac{i}{4}\right) \gamma,\
\rho_{10,23}=\left(\frac{1}{8}-\frac{i}{16}\right) \gamma,\
\rho_{10,24}=\left(-\frac{1}{16}-\frac{3 i}{16}\right) \gamma,\
\rho_{10,25}=\left(-\frac{5}{16}-\frac{i}{16}\right) \gamma\\
\rho_{10,26}&=-\frac{1}{16} (i \gamma ),\
\rho_{10,27}=\frac{i \gamma }{4},\
\rho_{10,28}=\left(\frac{1}{8}-\frac{i}{16}\right) \gamma,\
\rho_{10,29}=\left(-\frac{1}{16}-\frac{i}{16}\right) \gamma,\
\rho_{10,30}=\frac{3 i \gamma }{16}\\
\rho_{10,31}&=\frac{\gamma }{4},\
\rho_{11,12}=\left(-\frac{1}{8}-\frac{5 i}{16}\right) \gamma,\
\rho_{11,13}=-\frac{1}{8} (i \gamma ),\
\rho_{11,14}=-\frac{\gamma }{16},\
\rho_{11,15}=0\\
\rho_{11,16}&=\left(\frac{1}{16}+\frac{3 i}{16}\right) \gamma,\
\rho_{11,17}=0,\
\rho_{11,18}=\left(-\frac{1}{16}+\frac{i}{4}\right) \gamma,\
\rho_{11,19}=\left(\frac{1}{16}-\frac{i}{8}\right) \gamma,\
\rho_{11,20}=\left(-\frac{1}{16}+\frac{i}{16}\right) \gamma\\
\rho_{11,21}&=\left(\frac{1}{8}+\frac{i}{16}\right) \gamma,\
\rho_{11,22}=-\frac{1}{16} (3 i \gamma ),\
\rho_{11,23}=\left(\frac{1}{16}+\frac{i}{8}\right) \gamma,\
\rho_{11,24}=\left(-\frac{1}{16}+\frac{i}{16}\right) \gamma,\
\rho_{11,25}=\left(\frac{1}{16}+\frac{i}{16}\right) \gamma\\
\rho_{11,26}&=\left(\frac{1}{8}-\frac{i}{8}\right) \gamma,\
\rho_{11,27}=\left(-\frac{1}{4}+\frac{i}{16}\right) \gamma,\
\rho_{11,28}=\left(-\frac{1}{16}-\frac{i}{16}\right) \gamma,\
\rho_{11,29}=\left(\frac{1}{8}+\frac{i}{16}\right) \gamma,\
\rho_{11,30}=-\frac{1}{8} (i \gamma )\\
\rho_{11,31}&=\left(\frac{1}{8}-\frac{i}{16}\right) \gamma,\
\rho_{12,13}=\left(-\frac{1}{16}+\frac{i}{16}\right) \gamma,\
\rho_{12,14}=0,\
\rho_{12,15}=\left(\frac{3}{16}+\frac{i}{8}\right) \gamma,\
\rho_{12,16}=\frac{\gamma }{16}\\
\rho_{12,17}&=\left(-\frac{1}{8}-\frac{i}{16}\right) \gamma,\
\rho_{12,18}=-\frac{1}{16} (3 i \gamma ),\
\rho_{12,19}=\left(-\frac{3}{16}-\frac{i}{16}\right) \gamma,\
\rho_{12,20}=\left(-\frac{3}{16}-\frac{i}{8}\right) \gamma,\
\rho_{12,21}=-\frac{1}{16} (3 \gamma )\\
\rho_{12,22}&=\left(\frac{1}{8}+\frac{i}{8}\right) \gamma,\
\rho_{12,23}=\left(-\frac{1}{16}-\frac{3 i}{16}\right) \gamma,\
\rho_{12,24}=\left(-\frac{1}{4}-\frac{i}{16}\right) \gamma,\
\rho_{12,25}=\left(-\frac{1}{8}+\frac{i}{8}\right) \gamma,\
\rho_{12,26}=\left(-\frac{1}{8}+\frac{3 i}{16}\right) \gamma\\
\rho_{12,27}&=\left(\frac{1}{16}+\frac{i}{16}\right) \gamma,\
\rho_{12,28}=\frac{i \gamma }{16},\
\rho_{12,29}=-\frac{\gamma }{8},\
\rho_{12,30}=\left(\frac{1}{16}-\frac{i}{16}\right) \gamma,\
\rho_{12,31}=\left(-\frac{1}{16}+\frac{i}{16}\right) \gamma\\
\rho_{13,14}&=\left(\frac{1}{16}+\frac{i}{8}\right) \gamma,\
\rho_{13,15}=0,\
\rho_{13,16}=\left(-\frac{1}{4}-\frac{3 i}{16}\right) \gamma,\
\rho_{13,17}=\left(\frac{1}{16}+\frac{i}{8}\right) \gamma,\
\rho_{13,18}=\left(-\frac{3}{16}+\frac{i}{16}\right) \gamma\\
\rho_{13,19}&=-\frac{1}{16} (i \gamma ),\
\rho_{13,20}=-\frac{\gamma }{16},\
\rho_{13,21}=\left(\frac{1}{16}-\frac{i}{8}\right) \gamma,\
\rho_{13,22}=\left(\frac{3}{16}-\frac{3 i}{16}\right) \gamma,\
\rho_{13,23}=\frac{3 i \gamma }{8}\\
\rho_{13,24}&=-\frac{\gamma }{8},\
\rho_{13,25}=\left(-\frac{1}{8}+\frac{3 i}{16}\right) \gamma,\
\rho_{13,26}=\left(\frac{1}{16}+\frac{i}{16}\right) \gamma,\
\rho_{13,27}=\left(\frac{1}{8}+\frac{i}{16}\right) \gamma,\
\rho_{13,28}=\left(-\frac{1}{4}-\frac{i}{8}\right) \gamma\\
\rho_{13,29}&=-\frac{1}{16} (i \gamma ),\
\rho_{13,30}=\left(-\frac{1}{16}-\frac{i}{16}\right) \gamma,\
\rho_{13,31}=\left(\frac{3}{16}-\frac{i}{16}\right) \gamma,\
\rho_{14,15}=\left(-\frac{1}{16}-\frac{i}{16}\right) \gamma,\
\rho_{14,16}=\left(\frac{1}{8}-\frac{i}{16}\right) \gamma\\
\rho_{14,17}&=\left(-\frac{1}{16}-\frac{i}{16}\right) \gamma,\
\rho_{14,18}=\left(\frac{5}{16}-\frac{i}{4}\right) \gamma,\
\rho_{14,19}=\left(\frac{1}{8}+\frac{i}{16}\right) \gamma,\
\rho_{14,20}=\left(-\frac{1}{8}+\frac{i}{8}\right) \gamma,\
\rho_{14,21}=\left(-\frac{1}{16}+\frac{3 i}{16}\right) \gamma\\
\rho_{14,22}&=\left(\frac{1}{16}-\frac{i}{8}\right) \gamma,\
\rho_{14,23}=\left(\frac{1}{16}-\frac{i}{8}\right) \gamma,\
\rho_{14,24}=\frac{3 i \gamma }{16},\
\rho_{14,25}=\left(-\frac{1}{16}-\frac{5 i}{16}\right) \gamma,\
\rho_{14,26}=\left(-\frac{1}{8}+\frac{i}{16}\right) \gamma\\
\rho_{14,27}&=\left(-\frac{1}{8}-\frac{i}{8}\right) \gamma,\
\rho_{14,28}=\left(-\frac{1}{16}+\frac{i}{16}\right) \gamma,\
\rho_{14,29}=\left(-\frac{3}{16}-\frac{i}{16}\right) \gamma,\
\rho_{14,30}=\left(-\frac{1}{4}+\frac{i}{16}\right) \gamma,\
\rho_{14,31}=-\frac{1}{8} (i \gamma )\\
\rho_{15,16}&=\left(\frac{3}{16}+\frac{i}{16}\right) \gamma,\
\rho_{15,17}=\left(\frac{1}{8}+\frac{i}{16}\right) \gamma,\
\rho_{15,18}=-\frac{1}{16} (i \gamma ),\
\rho_{15,19}=\left(-\frac{3}{16}+\frac{i}{8}\right) \gamma,\
\rho_{15,20}=\left(\frac{3}{16}-\frac{i}{16}\right) \gamma\\
\rho_{15,21}&=-\frac{1}{8} (i \gamma ),\
\rho_{15,22}=\left(-\frac{1}{16}-\frac{i}{8}\right) \gamma,\
\rho_{15,23}=\left(-\frac{3}{16}+\frac{i}{8}\right) \gamma,\
\rho_{15,24}=\left(-\frac{1}{16}+\frac{3 i}{16}\right) \gamma,\
\rho_{15,25}=\left(\frac{1}{4}+\frac{i}{16}\right) \gamma\\
\rho_{15,26}&=-\frac{\gamma }{8},\
\rho_{15,27}=\left(\frac{1}{4}+\frac{i}{16}\right) \gamma,\
\rho_{15,28}=\left(\frac{1}{16}-\frac{3 i}{16}\right) \gamma,\
\rho_{15,29}=\left(\frac{1}{16}+\frac{i}{16}\right) \gamma,\
\rho_{15,30}=-\frac{\gamma }{8}\\
\rho_{15,31}&=-\frac{1}{16} (i \gamma ),\
\rho_{16,17}=\left(\frac{1}{16}-\frac{i}{16}\right) \gamma,\
\rho_{16,18}=\left(\frac{1}{4}-\frac{i}{4}\right) \gamma,\
\rho_{16,19}=\left(\frac{1}{16}-\frac{i}{4}\right) \gamma,\
\rho_{16,20}=-\frac{1}{8} (i \gamma )\\
\rho_{16,21}&=\frac{\gamma }{16},\
\rho_{16,22}=\left(-\frac{1}{4}+\frac{i}{8}\right) \gamma,\
\rho_{16,23}=\left(\frac{1}{8}+\frac{i}{16}\right) \gamma,\
\rho_{16,24}=\frac{3 \gamma }{16},\
\rho_{16,25}=-\frac{\gamma }{16}\\
\rho_{16,26}&=\left(\frac{1}{8}-\frac{i}{16}\right) \gamma,\
\rho_{16,27}=\left(-\frac{1}{8}-\frac{i}{16}\right) \gamma,\
\rho_{16,28}=\left(\frac{1}{16}+\frac{i}{16}\right) \gamma,\
\rho_{16,29}=0,\
\rho_{16,30}=\left(-\frac{1}{8}-\frac{i}{8}\right) \gamma
\end{align*}
\begin{align*}
\rho_{16,31}&=\left(\frac{1}{16}+\frac{i}{4}\right) \gamma,\
\rho_{17,18}=-\frac{\gamma }{16},\
\rho_{17,19}=\frac{i \gamma }{4},\
\rho_{17,20}=\frac{\gamma }{16},\
\rho_{17,21}=\left(\frac{1}{8}-\frac{i}{8}\right) \gamma\\
\rho_{17,22}&=\left(-\frac{1}{8}+\frac{i}{16}\right) \gamma,\
\rho_{17,23}=\left(\frac{1}{4}+\frac{i}{4}\right) \gamma,\
\rho_{17,24}=\frac{\gamma }{16},\
\rho_{17,25}=-\frac{1}{16} (3 \gamma ),\
\rho_{17,26}=-\frac{1}{16} (i \gamma )\\
\rho_{17,27}&=\left(\frac{1}{8}+\frac{i}{16}\right) \gamma,\
\rho_{17,28}=\frac{\gamma }{4},\
\rho_{17,29}=\left(\frac{1}{16}-\frac{3 i}{16}\right) \gamma,\
\rho_{17,30}=\frac{\gamma }{16},\
\rho_{17,31}=\left(\frac{1}{4}-\frac{i}{8}\right) \gamma\\
\rho_{18,19}&=\left(\frac{1}{16}-\frac{3 i}{16}\right) \gamma,\
\rho_{18,20}=-\frac{\gamma }{4},\
\rho_{18,21}=\left(\frac{1}{8}+\frac{i}{16}\right) \gamma,\
\rho_{18,22}=-\frac{\gamma }{8},\
\rho_{18,23}=\frac{3 \gamma }{16}\\
\rho_{18,24}&=\left(-\frac{1}{8}+\frac{3 i}{16}\right) \gamma,\
\rho_{18,25}=-\frac{1}{16} (i \gamma ),\
\rho_{18,26}=\left(\frac{1}{16}-\frac{3 i}{8}\right) \gamma,\
\rho_{18,27}=\frac{\gamma }{16},\
\rho_{18,28}=\frac{i \gamma }{4}\\
\rho_{18,29}&=\left(\frac{1}{16}-\frac{i}{8}\right) \gamma,\
\rho_{18,30}=\left(-\frac{3}{16}+\frac{i}{16}\right) \gamma,\
\rho_{18,31}=\left(-\frac{1}{8}-\frac{i}{8}\right) \gamma,\
\rho_{19,20}=\left(-\frac{1}{8}+\frac{i}{16}\right) \gamma,\
\rho_{19,21}=\left(-\frac{1}{4}-\frac{i}{8}\right) \gamma\\
\rho_{19,22}&=\left(-\frac{1}{16}-\frac{i}{2}\right) \gamma,\
\rho_{19,23}=0,\
\rho_{19,24}=\left(\frac{1}{8}-\frac{i}{16}\right) \gamma,\
\rho_{19,25}=\left(-\frac{1}{8}-\frac{3 i}{16}\right) \gamma,\
\rho_{19,26}=\frac{3 \gamma }{16}\\
\rho_{19,27}&=\left(-\frac{1}{16}+\frac{i}{8}\right) \gamma,\
\rho_{19,28}=\left(\frac{1}{16}-\frac{i}{8}\right) \gamma,\
\rho_{19,29}=\frac{\gamma }{8},\
\rho_{19,30}=\left(\frac{1}{8}-\frac{i}{8}\right) \gamma,\
\rho_{19,31}=\left(-\frac{3}{16}-\frac{3 i}{16}\right) \gamma\\
\rho_{20,21}&=\left(\frac{1}{16}-\frac{i}{16}\right) \gamma,\
\rho_{20,22}=0,\
\rho_{20,23}=-\frac{1}{16} (5 \gamma ),\
\rho_{20,24}=\left(-\frac{1}{16}-\frac{i}{16}\right) \gamma,\
\rho_{20,25}=-\frac{1}{8} (i \gamma )\\
\rho_{20,26}&=-\frac{\gamma }{8},\
\rho_{20,27}=\left(\frac{3}{16}-\frac{i}{8}\right) \gamma,\
\rho_{20,28}=\frac{\gamma }{16},\
\rho_{20,29}=-\frac{1}{16} (3 \gamma ),\
\rho_{20,30}=-\frac{1}{16} (i \gamma )\\
\rho_{20,31}&=\left(-\frac{1}{8}+\frac{i}{16}\right) \gamma,\
\rho_{21,22}=\frac{\gamma }{16},\
\rho_{21,23}=0,\
\rho_{21,24}=\frac{i \gamma }{8},\
\rho_{21,25}=\left(-\frac{1}{16}-\frac{i}{16}\right) \gamma\\
\rho_{21,26}&=\left(-\frac{1}{16}-\frac{i}{8}\right) \gamma,\
\rho_{21,27}=\frac{i \gamma }{4},\
\rho_{21,28}=-\frac{\gamma }{16},\
\rho_{21,29}=\frac{3 \gamma }{16},\
\rho_{21,30}=\frac{i \gamma }{16}\\
\rho_{21,31}&=-\frac{1}{16} (3 i \gamma ),\
\rho_{22,23}=\left(\frac{1}{16}+\frac{i}{16}\right) \gamma,\
\rho_{22,24}=\frac{i \gamma }{8},\
\rho_{22,25}=-\frac{\gamma }{16},\
\rho_{22,26}=\left(-\frac{1}{16}-\frac{i}{16}\right) \gamma\\
\rho_{22,27}&=\frac{\gamma }{8},\
\rho_{22,28}=-\frac{1}{16} (i \gamma ),\
\rho_{22,29}=\left(-\frac{1}{4}+\frac{i}{16}\right) \gamma,\
\rho_{22,30}=\left(\frac{3}{16}-\frac{i}{8}\right) \gamma,\
\rho_{22,31}=\left(\frac{3}{16}-\frac{i}{4}\right) \gamma\\
\rho_{23,24}&=\left(\frac{3}{16}-\frac{i}{4}\right) \gamma,\
\rho_{23,25}=\left(-\frac{1}{8}+\frac{i}{8}\right) \gamma,\
\rho_{23,26}=\left(\frac{1}{8}-\frac{i}{4}\right) \gamma,\
\rho_{23,27}=\left(-\frac{1}{16}-\frac{i}{16}\right) \gamma,\
\rho_{23,28}=\left(-\frac{1}{8}+\frac{i}{16}\right) \gamma\\
\rho_{23,29}&=-\frac{1}{16} (3 i \gamma ),\
\rho_{23,30}=\left(-\frac{3}{16}+\frac{i}{4}\right) \gamma,\
\rho_{23,31}=\left(\frac{1}{16}-\frac{i}{8}\right) \gamma,\
\rho_{24,25}=\left(-\frac{1}{16}-\frac{i}{16}\right) \gamma,\
\rho_{24,26}=0\\
\rho_{24,27}&=\left(-\frac{1}{16}-\frac{i}{4}\right) \gamma,\
\rho_{24,28}=0,\
\rho_{24,29}=\frac{\gamma }{16},\
\rho_{24,30}=0,\
\rho_{24,31}=\left(-\frac{1}{8}+\frac{i}{16}\right) \gamma\\
\rho_{25,26}&=-\frac{1}{16} (3 \gamma ),\
\rho_{25,27}=-\frac{\gamma }{4},\
\rho_{25,28}=\frac{\gamma }{16},\
\rho_{25,29}=\left(\frac{1}{8}+\frac{i}{4}\right) \gamma,\
\rho_{25,30}=\left(\frac{1}{8}+\frac{i}{16}\right) \gamma\\
\rho_{25,31}&=-\frac{1}{8} (i \gamma ),\
\rho_{26,27}=\left(\frac{3}{16}-\frac{3 i}{16}\right) \gamma,\
\rho_{26,28}=\frac{i \gamma }{8},\
\rho_{26,29}=\left(\frac{1}{8}+\frac{i}{16}\right) \gamma,\
\rho_{26,30}=\left(-\frac{1}{8}+\frac{i}{8}\right) \gamma\\
\rho_{26,31}&=\frac{3 \gamma }{16},\
\rho_{27,28}=\left(-\frac{1}{8}+\frac{i}{16}\right) \gamma,\
\rho_{27,29}=-\frac{1}{4} (i \gamma ),\
\rho_{27,30}=-\frac{\gamma }{16},\
\rho_{27,31}=-\frac{1}{8} (i \gamma )\\
\rho_{28,29}&=\left(-\frac{1}{16}+\frac{3 i}{16}\right) \gamma,\
\rho_{28,30}=0,\
\rho_{28,31}=\frac{\gamma }{16},\
\rho_{29,30}=-\frac{\gamma }{16},\
\rho_{29,31}=0,\ \rho_{30,31}=\left(-\frac{5}{16}-\frac{3 i}{16}\right) \gamma
\end{align*}
Again, from hermicity the lower-diagonal terms follow.

\end{document}